Generative AI Purpose-built for Social and Mental Health: A Real-World Pilot

Thomas D. Hull,[1] Lizhe Zhang,[1] Caitlin A. Stamatis[1], Patricia A. Arean[2] & Matteo Malgaroli[3*]

[1] Slingshot AI
[2] CREATIV Lab, Department of Psychiatry and Behavioral Sciences, University of Washington
[3] Department of Psychiatry, NYU School of Medicine

*Corresponding author: Matteo Malgaroli, PhD, 1 Park Ave, 8th Floor, New York NY.
matteo.malgaroli@nyulangone.org

**Abstract**

Generative AI chatbots built for mental health could extend access to care, but evidence from real-world use is limited. We report a single-arm, naturalistic pilot of a foundation model trained for mental health, among 299 US adults with at least moderate depressive or anxiety symptoms who were followed for up to 12 months. Depression and anxiety symptoms fell by 10 weeks (Cohen's d 0.93 and 0.79), with loneliness, behavioral activation, and social interaction improving. Clinicians confirmed that automated safeguards were escalated appropriately. Participants fell into non-responding (57.2%), improving (37.1%) and rapidly improving (5.7%) trajectories. Early working alliance and greater engagement were associated with better outcomes. The AI deployed ten identified intervention families whose delivery varied with baseline anxiety and depression, with the overall ratio of clinical to non-clinical content increasing according to severity without survey information access. These findings support the feasibility of purpose-built AI for mental health.

## INTRODUCTION

Mental health challenges continue to be a leading cause of disability worldwide[1] with poor access to care leaving many individuals untreated[2-3] due to geographic remoteness, financial constraints, competing demands, shortages of practitioners, stigma, and mobility impairments[4-7]. These persistent inequities underscore the importance of innovative approaches to extend the reach of high-quality mental health services[8].

Digital interventions can improve accessibility and scalability while maintaining effectiveness[9-10]. While most work has focused on live video therapy, growing evidence shows that message-based formats deliver equivalent outcomes[11-14]. This has spurred interest in chatbot-delivered interventions to reduce availability barriers and lower costs[15-17], making AI-driven chatbots a promising next step[18]. Early chatbots were designed to emulate therapeutic dialogue and deliver evidence-based strategies within pre-specified parameters[19-20]. Next generation Large Language Model (LLMs) chatbots promise greater personalization and more natural conversations[16,21].

Preliminary studies suggest mental health chatbots may be acceptable to users and can reduce symptoms of depression and anxiety. However, most trials to date have been small, short, or limited to a specific population[22-24]. The first randomized controlled trial (RCT) of a chatbot trained on clinically relevant data demonstrated promising outcomes[16]. An important next step is to evaluate the feasibility of this approach for engagement and effectiveness in real-world use. Given the ubiquity and growing reliance on generative AI (AI), larger, longer real-world evaluations are needed to determine scalability and impact of mental health chatbots.

Here we report a longitudinal, naturalistic evaluation of an AI model trained on mental health data (Figure 1), following 299 adults with at least moderate depression or anxiety symptoms in a longitudinal real-world setting up to 12 months. We assess the model's safety through clinician review of automated guardrails and performance on adversarial benchmarks designed for mental health AI. We examine trajectories across clinical and social outcomes and identify associated characteristics. Finally we use LLM transcript analysis validated by clinicians to identify the clinical interventions the model delivers. Prioritizing external validity, practical relevance, and long-term follow-up, we investigate utilization and engagement as indicators of acceptability and dosage to inform pragmatic AI implementations[25].

## RESULTS

### *Sample characteristics*

Participants ranged in age from 18 to 73 years, with an average of 41.7 years (SD=11.9). The sample was predominantly female (82%), and less than half (49.4%) reported having completed a bachelor's degree or more. At baseline (n = 299), mean scores were 15.45 (SD = 4.95) for the PHQ-9, 13.06 (4.39) for the GAD-7, 11.95 (1.83) for UCLA Loneliness Scale and 15.64 (7.01) for the BADS. Full demographic details are presented in Table 1.

[Table 1 about here]

***Safety Guardrails***

Of the 7,493 total sessions, 76 (1.02%) were flagged by in-app safety classifiers as containing potential crisis content. These instances were reviewed by clinicians to ensure the model escalated participants to appropriate resources, with all confirmed as correctly escalated (i.e., true positives). A random sample of 1,200 sessions were also checked for crisis messages missed by the escalation classifier (i.e., false negatives) and none were found (see Figure 1 for model architecture).

[Figure 1 about here]

***Appropriateness and Safety Benchmarks Performance***

Model safety and appropriateness were further tested in simulated evaluation without its safety guardrails as per production user-facing system. In the clinical test scenarios, the model produced 100 generations for each of the ten test prompts, returning a 87.7% pass rate (95% CI [85.5%, 89.6%]), within the bounds of the pass rate reported for human therapists[26]. SIM-VAIL[27] benchmark scores can be seen in Figure 2. Ash scores are lower than the median of nine widely used frontier models under the primary judge (1.39 vs 4.48 baseline median) and an independent alternate-family judge (2.47 vs 5.47). Ash over-refusal remained at floor with therapeutic quality near ceiling.

[Figure 2 about here]

***Clinical Change***

At 10 weeks, depression was lower than at baseline (PHQ-9: Cohen's $d = 0.93$, 95% CI: 0.76-1.13) and anxiety was also lower (GAD-7: $d = 0.79$, 95% CI: 0.65-0.95; both $p < 0.001$). These improvements were maintained at 12 months (PHQ-9: $d = 0.98$, 95% CI: 0.79-1.22; GAD-7: $d = 0.76$, 95% CI: 0.61-0.93; $p < 0.001$). The number of participants deteriorating from baseline to 6 weeks was 4 of 217 (1.84%) for GAD-7 and 5 of 217 (2.30%) for PHQ-9, and 3 of 217 (1.38%) on both GAD-7 and PHQ-9 (See Supplementary Table 2 for more detail). No significant differences emerged from sensitivity analyses for PHQ-9 and GAD-7 missingness (see Supplementary Table 3). By 6 weeks, 135 (62.2%) participants reported progress on their goals, 73 (33.6%) achieved their goals completely, and 10 (4.6%) reported no progress.

Three longitudinal patterns emerged shared across depression, anxiety, social support, and behavioral activation from baseline to 10 week follow-up:  Non-responders (n=171, 57.2%), Improving (n=111, 37.1%), and Rapid improving (n=17, 5.7%), and. Model fit criteria are reported in the supplementary materials. Baseline symptom severity was comparable across classes, with differences driven by rate and synchrony of change (see Figure 3).

[Figure 3 about here]

*Therapeutic Alliance*

Week 6 alliance scores by item averaged 3.96 (SD=0.97) for the entire measure, 4.09 (SD=0.99) for Bond, 3.81 (SD=0.97) for Task, and 3.97 (SD=0.95) for Goal. TRI items suggested internalization of the model remained moderately characteristic at follow-up, with participants recalling specific statements (M=5.8, SD=2.5) and the model "coming to mind" during moments of distress (M=5.5, SD=2.1).

*Engagement*

Two hundred ninety-six (99%) participants held at least one session in weeks 1 and 2, 256 (86%) in weeks 3 and 4, 216 (72%) in weeks 5 and 6, 162 (54%) in weeks 7 through 10, 126 (42%) in weeks 11 to 26, and 106 (35%) in weeks 27 to 52. During the six-week intervention period, participants averaged 82.2 user messages per week (SD = 102.0, median 50.8) across 3.17 sessions per week (SD = 2.14). From week 7 through six months, participants averaged 12.1 messages per week (SD = 33.3, median 1.5) across 0.46 sessions per week (SD = 0.92), with 185 (62%) active at least once during the period. From six through twelve months, participants averaged 14.2 messages per week (SD = 60.9, median 0.0) across 0.48 sessions per week (SD = 1.68), with 106 (35%) active at least once.

*Predictors of Trajectories*

We used a multinomial logistic regression to identify characteristics predicting trajectory membership. Results indicated that higher Week 2 WAI scores predicted Improving and Rapid Improving trajectories (Table 1). Week 2 therapeutic alliance scores also significantly predicted improvement for Week 6 PHQ-9 ($\beta$=-0.16, SE=0.04, $r^2$=0.09, p<.001) and GAD-7 ($\beta$=-0.15, SE=0.03, $r^2$=0.11, p<.001).

Results from the mixed-effects analyses indicated that reductions in PHQ-9 scores were significantly associated with increased engagement for Improving (β = -1.30, 95% CI: -2.06, -0.54; p = .002) and Rapid Improving participants (β = -2.84, 95% CI: -4.88, -0.80; p = .007). Similarly, greater engagement was significantly associated with larger decreases in anxiety symptoms for Improving (β = -0.78, 95% CI: -1.50, -0.06; p = .034) and Rapid Improving participants (β = -3.10, 95% CI: -5.03, -1.16; p = .004). Full coefficients are reported in the Supplementary Materials.

[FIGURE 4 ABOUT HERE]

*Clinical Behavior of Model Characteristics*

Of 299 participants, 286 had conversation transcripts consented for technique coding and generated a median of 55 six-turn conversational segments (IQR 27-100) per two week treatment window.

Applied to 606 conversations from 50 participants withheld from all outcome analyses, 37,453 descriptions of distinct therapist moves were extracted, 98% of which had no vocabulary overlap suggesting they could not be discovered by keyword or lexicon methods alone. Iterative LLM summarisation and refinement over these descriptions converged on a stable category set across 15 refinement rounds while fit ratings rose from 92 to 94 and plateaued. The resulting taxonomy comprised ten clinical technique families plus a non-clinical/session-management category retained as a negative control (See Supplementary Materials for category detail).

Clinicians and classifiers completely agreed on 31 of the total 55 cases, with clinician majority (two of the three) confirming 41 of 44 positive classifier cases (93%, 95% CI 82–98%) plus all 11 negative cases, tracking the withheld classifier score strongly (Spearman ρ = 0.568–0.678, all p < 0.0001). Each standard-deviation increase in prior PHQ-9 predicted a higher clinical share of the subsequent two-week window (standardized $\beta = 0.26$, $p < .001$), with the same pattern for GAD-7 ($\beta = 0.20$, $p = .014$), while the share of non-clinical content decreased (PHQ-9: $\beta = -0.27$, $p < .001$, $q = .004$; GAD-7: $\beta = -0.21$, $p = .013$, $q = .046$).

Model intervention delivery rates varied substantially across techniques (Table 2), with empathic validation present in 97.1% (95% CI: 96.6, 97.5) of segments. Non-clinical session management appeared in 52.7% of segments (95% CI: 50.9, 54.6) and was excluded from further analysis given its organizational nature. Collaborative process was the most frequently delivered therapeutic intervention, present in 51.6% of segments (95% CI: 49.6, 53.6), followed by cognitive reappraisal (45.5%, 95% CI: [43.2, 47.8]), reflective observation and pattern identification (39.4%, 95% CI [37.0, 41.8]), and behavioral activation (31.5%, 95% CI: [29.5, 33.5]).  The proportion of interventions using experiential and metaphor-based  (17.8%, 95% CI: [16.2, 19.4]), psychoeducation (16.1%, 95% CI: [14.8, 17.5]), communication coaching (13.4%, 95% CI: [11.9, 14.9]), and support mobilization (7.8%, 95% CI: [6.9, 8.7]) approaches were less frequent, with grounding and emotion regulation (5.9%, 95% CI: [5.0, 6.7]) as the least observed intervention category.

[Table 2 about here]

Higher baseline depression was associated with greater delivery of support mobilization (OR=1.29, 95% CI [1.14, 1.47], $p<.001$), experiential and metaphor interventions (OR=1.22, 95% CI [1.11, 1.36], $p<.001$), and collaborative process (OR=1.18, 95% CI [1.09, 1.28], $p<.001$). Higher baseline anxiety was associated with greater delivery of grounding and emotion regulation (OR=1.25, 95% CI [1.08, 1.45], $p=.016$). Lower delivery of communication coaching with higher depression did not survive correction for multiple testing (OR=0.84, 95% CI [0.74, 0.97], $p=.053$).

## Discussion

This longitudinal real world pilot examined symptom trajectories among users of a purpose-built mental health generative AI model. Our study complements randomized studies of AI technology by examining real-world use of a mental health AI model, and by incorporating additional social functioning and engagement analyses to elucidate how therapeutic processes may differentially benefit users of AI-delivered mental health support.

We observed symptom reduction across measures of depression and anxiety, with deterioration rates far less than those reported for waitlist controls[28] and for up to 12 months, the longest use of an AI yet reported. Alongside symptom reduction, we also examined changes in social health and identified associated trajectories of improvements in social support and social interactions. The latter finding is notable, given concerns raised about the potential misuse of chatbots to substitute for in-person interactions[29]. Our findings align with digital interventions as relational bridges rather than substitutes[30], and suggests that AI models that draw attention to existing social resources, and that help build social skills, could have the potential to support social health.

Another feature of the study was the use of automatic safety guardrails. In light of concerns about the unpredictable or inappropriate behavior of LLMs in health contexts[26,31], a key finding was the apparent accuracy of the guardrails in detecting and escalating participant messages with potential risk. These findings are consistent with an evaluation of the model's performance on external benchmarks, which show lower rates of harmful responding relative to frontier models (Ash 6.2% vs frontier models 18.0-52.0%), and no missed escalations in response to suicide-related conversations in an audit of 20,000 real-world conversations[32]. While these preliminary results are encouraging, the exclusion of individuals with acute suicidality, psychosis, and substance use precludes generalization and warrants further careful investigation of the feasibility of AI for these high-risk populations.

Therapeutic alliance with the model was comparable to traditional psychotherapy[33] and higher than rule-based chatbots[34] and internet-based CBT[35], consistent with previous AI findings[16]. Additionally, model internalization was characteristic and comparable to reported psychotherapy scores,[36] warranting further exploration.

We explored potential mediators of differential improvements from working alliance and engagement. Higher alliance was associated with greater reductions in anxiety and depression, replicating psychotherapy findings[37]. Similarly, increased days and minutes of usage were associated with symptom reductions in improving patients. These initial findings echo psychotherapy research demonstrating the importance of these constructs across modalities and diagnostic groups[37-39], suggesting alliance and engagement heterogeneity as potential mechanisms in AI mental health support for future investigation.

We also examined categories of common clinical interventions deployed by the AI model and whether delivery varied with depression and anxiety baseline symptoms. While empathic validation was present in almost all interaction segments examined, other interventions tracked the symptomatology of participants despite the model having no access to symptoms data. Specifically, participants with more severe depression were more often encouraged to mobilize their social network, more often engaged through experiential work, and more often invited to increase agency over their treatment, whereas participants with more severe anxiety more frequently received grounding exercises consistent with mindfulness approaches[40]. These preliminary findings suggest that purpose-built wellbeing AI presents a sensitivity to participant functioning inferred through message content and is able to use it as a context to select intervention categories consistent with clinical guidelines[41]. Whether the model is responding to implicit signals in the conversation or to explicit symptom disclosure and whether this differential selection affects outcomes, will require further evaluation. Larger studies could examine whether interventions dynamically shift as symptoms change over the course of treatment, and whether delivery responds to symptom clusters and subtypes beyond overall severity[42].

Limitations include the lack of a control condition, which limits causal claims about the intervention, as improvements may reflect spontaneous remission or placebo effects. However, our findings are consistent with psychotherapy trials and exceed the 16% meta-analytic symptom improvement among therapy control groups[43], the placebo effect size of g=0.28 for digital interventions[44] and the 12.5% spontaneous remission rates reported for depressive symptoms[28]. Study exclusions for fraudulent activity were higher than human behavioral intervention studies[13], requiring confirmation in future work. However, participant exclusions for clinical criteria were half those reported for AI[16] and non-AI digital mental health trials[13].

Despite these limitations, this study provides the first real-world demonstration of feasibility for a clinically grounded and health promoting AI model for mental health. The study advances a new class of AI-based chatbots with the potential to expand access without compromising safety or mental health outcomes.

## Methods

### *Setting*

The study included data from users who utilized Ash, a foundation model for mental health (talktoash.com). Ash is accessible via Apple App and Google Play stores, and internet search. The system is powered by an AI model trained on mental health-relevant data (e.g., transcripts, clinical interviews, case studies) sourced with permission by those receiving care at a variety of behavioral health settings throughout the US and are de-identified at the source before transmission.

Onboarding consisted of an opt-in consent procedure allowing use of de-identified data, followed by brief demographic questions. Participants then completed baseline measures designed to evaluate their current state (described below). After completing these measures, participants were introduced to Ash. All observations included in this study were collected between May 15, 2025 and September 15, 2026, as part of Slingshot AI's ongoing quality assurance and program management processes. Study procedures were approved by the Institutional Review Board at New York University School of Medicine (i26-00386).

### *Participants*

Participants responding to ads on Meta were recruited to participate through the app platform. Study eligibility was determined based on completion of standardized self-report measures during onboarding and availability of at least one additional data point. Inclusion criteria consisted of: 1) English-speaking users located in the United States, 2) age between 18 and 85, 3) reliable internet or smartphone access, 4) baseline scores of 10 or higher on either the PHQ-9 or GAD-7, and 5) at least one follow-up assessment during the study window. Exclusion criteria consisted of self-reported or flagged risk factors indicating a need for a higher level of care, including any schizophrenia spectrum or other psychotic disorder, severe substance or alcohol use disorder, medical or neurological conditions that would better account for symptoms, and active suicidal thoughts or behaviors sufficient to warrant "Yes" responses on items 3 to 6 of the Columbia Suicide Severity Rating Scale (C-SSRS), indicating risk that required immediate referral or emergency intervention.

A total of 1,808 user records were reviewed. Of these, 563 were excluded as potentially fraudulent, 344 declined participation, and 56 experienced technical issues. Of the remaining 845 individuals, 546 did not meet clinical inclusion criteria. CONSORT flow diagram and further information about recruitment are reported in the supplementary materials. The final sample consisted of 299 participants who met all inclusion criteria, had completed baseline assessments, and provided sufficient follow-up data to be analyzed.

### *AI model and Intervention*

The intervention is a mental health AI model delivered through text- and voice-based interaction for mobile devices. After each conversation, users receive automatically generated summaries, well-being suggestions, and personalized conversation starters intended to scaffold future engagement and reinforce momentum.

The AI is a domain-adapted large language model designed for mental health support. It is developed by training a general-purpose pretrained language model on over nine hundred thousand hours of anonymized mental health-relevant transcripts, a private corpus representing one of the largest scale datasets of its kind. The training corpus includes diverse therapeutic approaches such as psychodynamic therapy, Dialectical Behavior Therapy, Acceptance and Commitment Therapy, second-wave Cognitive Behavioral Therapy, and Motivational Interviewing, enabling the model to represent several evidence-based practice interventions to allow for user preference and personalization. Development further includes supervised fine-tuning on therapy-relevant conversational data and policy-based alignment, informed by clinician-generated annotations and human feedback on model outputs, to shape desired response behavior and reduce undesired responses. As of September 2026, Ash has collected 15.5k human annotations and has 18 active safety guardrails.

***Assessments***

Participants were assessed at baseline, Week 2, Week 4, and Week 6 as the intervention period, with follow-ups at Week 10, Month 6, and Month 12. Assessments were presented as an integral part of the experience, supporting reflection, progress tracking, and individualized goal setting.

The Patient Health Questionnaire (PHQ-9[45]) assessed severity of depressive symptoms. The Generalized Anxiety Disorder Scale (GAD-7[46]) was used to measure symptoms of anxiety. The Behavioral Activation for Depression Scale Short Form (BADS[47]) was used to assess activation and avoidance behaviors relevant to depression. The UCLA Loneliness Scale 4-item version[48,49] measured loneliness. The Multidimensional Scale of Perceived Social Support (MSPSS[50]) measured support from family, friends, and significant others. Social interaction subscale items from the Duke Social Support Index (DSSI[51]) captured social connectedness.

The Goal Attainment Scaling procedure[52] was adapted to capture individualized progress toward participant-defined goals. At baseline, participants were asked to identify their top three personal goals and to rate each goal on importance and difficulty on a 4-point scale. At follow-up assessments, participants rated the status of each goal.

The Working Alliance Inventory – Short Report (WAI-SR[53]) was administered to measure participants' perceptions of the therapeutic alliance with the AI model. We adapted the WAI's for this setting (e.g., "I believe Ash understands what I am trying to accomplish."), with excellent internal consistency (Cronbach's alpha=.96). The adapted Therapist Representation Inventory (TRI[36]) measured the extent to which participants internalized the GAI model's guidance.

***Safety and Appropriateness: Guardrails and Benchmarks***

The model is equipped with mental health specific safety guardrails and classifiers to ensure responsible practice. Each session flagged by the model's safety guardrails was reviewed by a clinician to verify that information for accessing human support, crisis lines, or emergency services were provided by the model in each case.

To further assess the presence of certain model shortcoming to core therapeutic principles (e.g., "Don't enable suicidal ideation" and "Don't reinforce hallucinations") we used an evaluation by Moore et al.[26], as well as SIM-VAIL a public, multi-turn benchmark for mental health[27]. The first comprises ten documented clinical scenarios (delusions, hallucinations, suicidal ideation, obsessive-compulsive disorder, and mania) to elicit responses and determine their consistency with clinical standards. For each scenario, we generated 100 responses using default model parameters. SIM-VAIL is a clinically validated adversarial audit framework in which a frontier model role-plays a user with a defined psychiatric vulnerability (psychosis, depression, obsessive-compulsive disorder, mania/hypomania, or insecure attachment) paired with a dominant conversational intent (validation of a maladaptive belief, permission for a risky action, reassurance and avoidance, dependence, trivialisation, or glorification), engaging the target chatbot over up to ten conversational turns. All 30 vulnerability × intent conditions were run in triplicate (n = 90 conversations) using default model parameters, and scored with an independent LLM judge across 39 behavioural dimensions.

### *Data Analytic Strategy*

*Power Analysis*

With N=299 and 80% power, the minimal detectable correlation was $r \approx 0.16$ ($R^2 \approx 0.026$), increasing to 0.19 at 90% power. Under 25% attrition (N=225), detectable correlations were $r \approx .19$ (80% power) and 0.21 (90% power). Under 45% attrition (N=165), detectable correlations increased to $r \approx 0.23$ (80% power) and 0.26 (90% power). These correspond to small effects explaining 4%-7% of variance at the highest thresholds. For within-person change, with N=299, the minimal detectable effect was Cohen's $d \approx$ 0.16 at 80% power and $d \approx 0.19$ at 90% power. Under 25% attrition (N=225), detectable d values were $\approx$ 0.19 (80% power) and 0.22 (90% power). Under 45% attrition (N=165), thresholds were 0.23 (80%) and 0.26 (90%). These estimations indicate that the study is powered to detect small associations ($r \approx$ 0.16-0.26) and small within-person changes ($d \approx$ 0.16-0.26) across plausible attrition scenarios.

*Clinical Change*

We calculated Cohen's d effect sizes at each assessment time point to quantify the magnitude of change in scores over time. We conducted Latent Growth Mixture Modeling (LGMM) using MPLUS 8 to identify trajectories across four clinical outcomes: depression (PHQ-9), anxiety (GAD-7), social support (DSSI), and behavioral activation (BADS). We used a parallel process LGMM to estimate heterogeneous change patterns across all four outcomes at once. The LGMM trajectories were estimated with full information maximum likelihood under missing not at random (MNAR) assumptions using the Muthén-Roy model to condition trajectory estimations on missing data patterns[54]. After determining the trajectories, we conducted a multinomial logistic regression to identify predictors among baseline demographics (age, gender, race, education, and income), concurrent clinical treatments (psychotropic medication use, psychotherapy), and initial therapeutic alliance. Missing values for predictors were imputed using multiple chained equations with random forest conditional models.

*Engagement*

Deidentified usage from the model platform was collected throughout the six week period: distinct days the platform was accessed, minutes of use, and total words exchanged by the participant. For each two week assessment window, we derived a composite engagement score defined as the first principal component of these three quantitative engagement metrics (see Supplementary Materials). We then examined the relationship between engagement and symptom improvement, as measured by changes in PHQ-9 and GAD-7 scores within each two-week assessment window (weeks 0-2, weeks 2-4, and weeks 4-6). Mixed-effects models were estimated by maximum likelihood using the package lme4[55] as follows:

$$\Delta Score_{ij} = \beta_0 + \beta_1 Engagement_{ij} + \beta 2 Trajectory_i + \beta 3(Engagement_{ij} \times Trajectory_i) + \beta_4 Window_j + u_i + \varepsilon_{ij}$$

where $\Delta Score_{ij}$ is the PHQ-9 and GAD-7 score change for participant *i* within the two-week interval *j*; $\beta_0$ is the intercept, $\beta_1$ is the effect of engagement in the same time window; $\beta_2$ captures the mean difference in symptom score change for the LGMM trajectory; $\beta_3$ is the interaction coefficient between engagement and trajectory; $\beta_4$ adjusts for the two-week assessment window; $u_i$ is a random intercept to account for between-subject variability in baseline change, and $\varepsilon_{ij}$ is the residual error for each observation. Separate models were fit to capture trajectory specific interaction slopes for PHQ-9 and GAD-7 scores, and P-values were adjusted using Holm-Bonferroni correction for multiple testing.

*Clinical Behavior of Model*

In-session model behavior was measured by inducing a taxonomy of clinical interventions directly from transcripts using TnT-LLM, a two-phase framework in which a large language model first generates and iteratively refines a label taxonomy from a corpus sample, then acts as an annotator to produce training data for lightweight classifiers deployable at scale[56,57]. Induction was performed on a 50-participant holdout and transcripts were then divided into six-turn windows and scored by intervention category logistic classifiers calibrated by isotonic regression against a stratified gold-standard sample and then blind reviewed by three clinical psychologists each independently judging 55 stratified, length-matched transcript excerpts for the presence of a single named technique (detail in the Supplementary Methods).

Conversations in the six weeks after onboarding were divided into two-week windows as per assessment schedule to count segments of at least six consecutive talk turns. Therapeutic interventions administered by the model were then identified and counted once per segment, with co-occurrences possible, based on the taxonomy of interventions identified by TnT-LLM.

For each participant, we counted the number of segments containing each intervention out of the total segments of that participant. We then performed exploratory analyses associating the proportion of segments containing an intervention with baseline depression and anxiety severity. Specifically, we fitted a beta-binomial mixed-effects model with a logit link for all interventions estimated by maximum likelihood using glmmTMB[58] as follows:

$$\text{logit}(Intervention_{ik}) = \beta_0 + \beta_1 PHQ\text{-}9_i + \beta_2 GAD\text{-}7_i + u_i$$

where $Intervention_{ik}$ is the proportion of segments of participant *i* containing intervention *k*; $\beta_0$ is the intercept; $\beta_1$ is the effect of baseline standardized PHQ-9 score and $\beta_2$ of baseline standardized GAD-7 score; and $u_i$ is a random intercept accounting for between participant variability. The intercept, slopes, and the dispersion parameters were estimated separately for each intervention $k$. Significance was adjusted for multiple testing using the Benjamini-Hochberg procedure.

**CRediT authorship contribution statement**

**Thomas D. Hull**: Conceptualization, Funding acquisition, Investigation, Project administration, Resources, Visualization, Writing - original draft, Writing - review and editing. **Lizhe Zhang**: Data curation, Investigation, Methodology, Software, Validation, Writing - review and editing. **Caitlin A. Stamatis**: Conceptualization, Investigation, Project administration, Writing - review and editing. **Patricia A. Arean**: Conceptualization, Methodology, Supervision, Writing - original draft, Writing - review and editing. **Matteo Malgaroli**: Conceptualization, Data curation, Formal analysis, Funding acquisition, Methodology, Resources, Supervision, Validation, Visualization, Writing - original draft, Writing - review and editing.

**Competing Interests**

TDH, LZ, and CAS are current or former employees of the company building the model and declare no non-financial competing interests. PAA holds advisor shares in Slingshot and declares no non-financial competing interests. MM has received a research grant from Slingshot AI administered through NYU Grossman School of Medicine.

**Role of Funding**

M.M. discloses support for the research of this work from the National Institute of Mental Health [K23MH134068], the National Center for Complementary and Integrative Health [R34AT012943] and Slingshot AI. T.D.H., L.Z. and C.A.S. disclose support for the research of this work from Slingshot AI, their current or former employer. P.A.A. declares no relevant funding. [The funders had no role in the analysis, interpretation of the data or decision to submit the manuscript for publication.] The content is solely the responsibility of the authors and does not necessarily represent the official views of the National Institutes of Health.

**Availability of Data and materials**

Quantitative data are available by reasonable request and a data use agreement. Transcript data are not available at this time due to the sample size and sensitivity of the data.

**Code Availability**

Code for the analyses in this study has been made available and can be found on the Open Science Foundation repository: https://osf.io/a7m6e/overview?view_only=bf4577f0998a428b856b7383f45376fe

## References

1. World Health Organization. Depression and Other Common Mental Disorders: Global Health Estimates (WHO, 2017).
2. Kazdin, A. E. & Rabbitt, S. M. Novel models for delivering mental health services and reducing the burdens of mental illness. Clin. Psychol. Sci. 1, 170–191 (2013).
3. Thornicroft, G. et al. Undertreatment of people with major depressive disorder in 21 countries. Br. J. Psychiatry 210, 119–124 (2017).
4. Andrade, L. H. et al. Barriers to mental health treatment: results from the WHO World Mental Health surveys. Psychol. Med. 44, 1303–1317 (2014).
5. Clement, S. et al. What is the impact of mental health-related stigma on help-seeking? A systematic review of quantitative and qualitative studies. Psychol. Med. 45, 11–27 (2015).
6. Mojtabai, R. et al. Barriers to mental health treatment: results from the National Comorbidity Survey Replication. Psychol. Med. 41, 1751–1761 (2011).
7. Wang, P. S. et al. Use of mental health services for anxiety, mood, and substance disorders in 17 countries in the WHO world mental health surveys. Lancet 370, 841–850 (2007).
8. Patel, V. et al. The Lancet Commission on global mental health and sustainable development. Lancet 392, 1553–1598 (2018).
9. Andersson, G., Cuijpers, P., Carlbring, P., Riper, H. & Hedman-Lagerlöf, E. Guided internet-based vs. face-to-face cognitive behavior therapy for psychiatric and somatic disorders: a systematic review and meta-analysis. World Psychiatry 13, 288–295 (2014).
10. Carlbring, P., Andersson, G., Cuijpers, P., Riper, H. & Hedman-Lagerlöf, E. Internet-based vs. face-to-face cognitive behavior therapy for psychiatric and somatic disorders: an updated systematic review and meta-analysis. Cogn. Behav. Ther. 47, 1–18 (2018).
11. Areán, P. A. et al. Randomized trial of the effectiveness of videoconferencing-based versus message-based psychotherapy on depression. Psychiatr. Serv. 75, 1184–1191 (2024).
12. Hull, T. D., Malgaroli, M., Connolly, P. S., Feuerstein, S. & Simon, N. M. Two-way messaging therapy for depression and anxiety: longitudinal response trajectories. BMC Psychiatry 20, 297 (2020).
13. Pullmann, M. D. et al. Message-based vs video-based psychotherapy for depression: a randomized clinical trial. JAMA Netw. Open 8, e2540065 (2025).
14. Song, J. et al. Comparing message-based psychotherapy to once-weekly, video-based psychotherapy for moderate depression: randomized controlled trial. J. Med. Internet Res. 25, e46052 (2023).
15. Abd-Alrazaq, A. A. et al. An overview of the features of chatbots in mental health: a scoping review. Int. J. Med. Inform. 132, 103978 (2019).
16. Heinz, M. V. et al. Randomized trial of a generative AI chatbot for mental health treatment. NEJM AI 2, AIoa2400802 (2025).
17. Linardon, J., Cuijpers, P., Carlbring, P., Messer, M. & Fuller-Tyszkiewicz, M. The efficacy of app-supported smartphone interventions for mental health problems: a meta-analysis of randomized controlled trials. World Psychiatry 18, 325–336 (2019).
18. Malgaroli, M. et al. Large language models for the mental health community: framework for translating code to care. Lancet Digit. Health 7, e282–e285 (2025).

19. Fitzpatrick, K. K., Darcy, A. & Vierhile, M. Delivering cognitive behavior therapy to young adults with symptoms of depression and anxiety using a fully automated conversational agent (Woebot): a randomized controlled trial. JMIR Ment. Health 4, e7785 (2017).
20. Inkster, B., Sarda, S. & Subramanian, V. An empathy-driven, conversational artificial intelligence agent (Wysa) for digital mental well-being: real-world data evaluation mixed-methods study. JMIR mHealth uHealth 6, e12106 (2018).
21. Ovsyannikova, D., de Mello, V. O. & Inzlicht, M. Third-party evaluators perceive AI as more compassionate than expert humans. Commun. Psychol. 3, 4 (2025).
22. Fulmer, R., Joerin, A., Gentile, B., Lakerink, L. & Rauws, M. Using psychological artificial intelligence (Tess) to relieve symptoms of depression and anxiety: randomized controlled trial. JMIR Ment. Health 5, e9782 (2018).
23. Reyes-Portillo, J. A. et al. Generative AI–powered mental wellness chatbot for college student mental wellness: open trial. JMIR Form. Res. 9, e71923 (2025).
24. Vaidyam, A. N., Wisniewski, H., Halamka, J. D., Kashavan, M. S. & Torous, J. B. Chatbots and conversational agents in mental health: a review of the psychiatric landscape. Can. J. Psychiatry 64, 456–464 (2019).
25. Mohr, D. C., Weingardt, K. R., Reddy, M. & Schueller, S. M. Three problems with current digital mental health research… and three things we can do about them. Psychiatr. Serv. 68, 427–429 (2017).
26. Moore, J. et al. Expressing stigma and inappropriate responses prevents LLMs from safely replacing mental health providers. In Proc. 2025 ACM Conf. Fairness, Accountability, and Transparency 599–627 (ACM, 2025).
27. Weilnhammer, V. et al. A clinically validated framework for auditing AI chatbot behavior in mental health interactions. Nat. Med. https://doi.org/10.1038/s41591-026-04577-2 (2026).
28. Cuijpers, P. et al. The effects of psychotherapies for depression on response, remission, reliable change, and deterioration: a meta-analysis. Acta Psychiatr. Scand. 144, 288–299 (2021).
29. Coghlan, S. et al. To chat or bot to chat: ethical issues with using chatbots in mental health. Digit. Health 9, 20552076231183542 (2023).
30. Baumel, A., Fleming, T. & Schueller, S. M. Digital micro interventions for behavioral and mental health gains: core components and conceptualization of digital micro intervention care. J. Med. Internet Res. 22, e20631 (2020).
31. McBain, R. K. et al. Evaluation of alignment between large language models and expert clinicians in suicide risk assessment. Psychiatr. Serv. 76, 944–950 (2025).
32. Stamatis, C. et al. Beyond simulations: what 20,000 real conversations reveal about mental health AI safety. Preprint at https://arxiv.org/abs/2601.17003 (2026).
33. Munder, T., Wilmers, F., Leonhart, R., Linster, H. W. & Barth, J. Working Alliance Inventory-Short Revised (WAI-SR): psychometric properties in outpatients and inpatients. Clin. Psychol. Psychother. 17, 231–239 (2010).
34. Darcy, A., Daniels, J., Salinger, D., Wicks, P. & Robinson, A. Evidence of human-level bonds established with a digital conversational agent: cross-sectional, retrospective observational study. JMIR Form. Res. 5, e27868 (2021).
35. Jasper, K. et al. The working alliance in a randomized controlled trial comparing Internet-based self-help and face-to-face cognitive behavior therapy for chronic tinnitus. Internet Interv. 1, 49–57 (2014).

36. Geller, J. & Farber, B. Factors influencing the process of internalization in psychotherapy. Psychother. Res. 3, 166–180 (1993).
37. Falkenström, F., Granström, F. & Holmqvist, R. Working alliance predicts psychotherapy outcome even while controlling for prior symptom improvement. Psychother. Res. 24, 146–159 (2014).
38. Burkhardt, H. A. et al. Behavioral activation and depression symptomatology: longitudinal assessment of linguistic indicators in text-based therapy sessions. J. Med. Internet Res. 23, e28244 (2021).
39. Wampold, B. E. & Imel, Z. E. The Great Psychotherapy Debate: The Evidence for What Makes Psychotherapy Work (Routledge, 2015).
40. Hoge, E. A. et al. Mindfulness-based stress reduction vs escitalopram for the treatment of adults with anxiety disorders: a randomized clinical trial. JAMA Psychiatry 80, 13–21 (2023).
41. Keynejad, R. C., Dua, T., Barbui, C. & Thornicroft, G. WHO Mental Health Gap Action Programme (mhGAP) intervention guide: a systematic review of evidence from low and middle-income countries. Evid. Based Ment. Health 21, 30–34 (2018).
42. Mermin, Z. et al. Anxiety and depression subtypes and their psychotherapy response: a network analysis of 33,675 patients. Psychol. Med. 56, e257 (2026).
43. Hosono, T., Tsutsumi, R., Niwa, Y. & Kondoh, M. Magnitude of the digital placebo effect and its moderators on generalized anxiety symptoms: systematic review and meta-analysis. J. Med. Internet Res. 27, e74905 (2025).
44. Mekonen, T. et al. What is the short-term remission rate for people with untreated depression? A systematic review and meta-analysis. J. Affect. Disord. 296, 17–25 (2022).
45. Kroenke, K., Spitzer, R. L. & Williams, J. B. The PHQ-9: validity of a brief depression severity measure. J. Gen. Intern. Med. 16, 606–613 (2001).
46. Spitzer, R. L., Kroenke, K., Williams, J. B. & Löwe, B. A brief measure for assessing generalized anxiety disorder: the GAD-7. Arch. Intern. Med. 166, 1092–1097 (2006).
47. Manos, R. C., Kanter, J. W. & Luo, W. The behavioral activation for depression scale–short form: development and validation. Behav. Ther. 42, 726–739 (2011).
48. Russell, D., Peplau, L. A. & Cutrona, C. E. The revised UCLA Loneliness Scale: concurrent and discriminant validity evidence. J. Pers. Soc. Psychol. 39, 472–480 (1980).
49. Hughes, M. E., Waite, L. J., Hawkley, L. C. & Cacioppo, J. T. A short scale for measuring loneliness in large surveys: results from two population-based studies. Res. Aging 26, 655–672 (2004).
50. Zimet, G. D., Dahlem, N. W., Zimet, S. G. & Farley, G. K. The Multidimensional Scale of Perceived Social Support. J. Pers. Assess. 52, 30–41 (1988).
51. Wardian, J., Robbins, D., Wolfersteig, W., Johnson, T. & Dustman, P. Validation of the DSSI-10 to measure social support in a general population. Res. Soc. Work Pract. 23, 100–106 (2013).
52. Kiresuk, T. J. & Sherman, R. E. Goal attainment scaling: a general method for evaluating comprehensive community mental health programs. Community Ment. Health J. 4, 443–453 (1968).
53. Hatcher, R. L. & Gillaspy, J. A. Development and validation of a revised short version of the Working Alliance Inventory. Psychother. Res. 16, 12–25 (2006).

54. Muthén, B., Asparouhov, T., Hunter, A. M. & Leuchter, A. F. Growth modeling with nonignorable dropout: alternative analyses of the STAR*D antidepressant trial. Psychol. Methods 16, 17–33 (2011).
55. Bates, D., Mächler, M., Bolker, B. & Walker, S. Fitting linear mixed-effects models using lme4. J. Stat. Softw. 67, 1–48 (2015).
56. Wan, M. et al. TnT-LLM: text mining at scale with large language models. In Proc. 30th ACM SIGKDD Conf. Knowledge Discovery and Data Mining 5836–5847 (ACM, 2024).
57. Lam, M. S., Teoh, J., Landay, J. A., Heer, J. & Bernstein, M. S. Concept induction: analyzing unstructured text with high-level concepts using LLooM. In Proc. 2024 CHI Conf. Human Factors in Computing Systems 1–28 (ACM, 2024).
58. McGillycuddy, M., Popovic, G., Bolker, B. M. & Warton, D. I. Parsimoniously fitting large multivariate random effects in glmmTMB. J. Stat. Softw. 112, 1–19 (2025).

**Figure 1.** Diagram of Ash architecture.

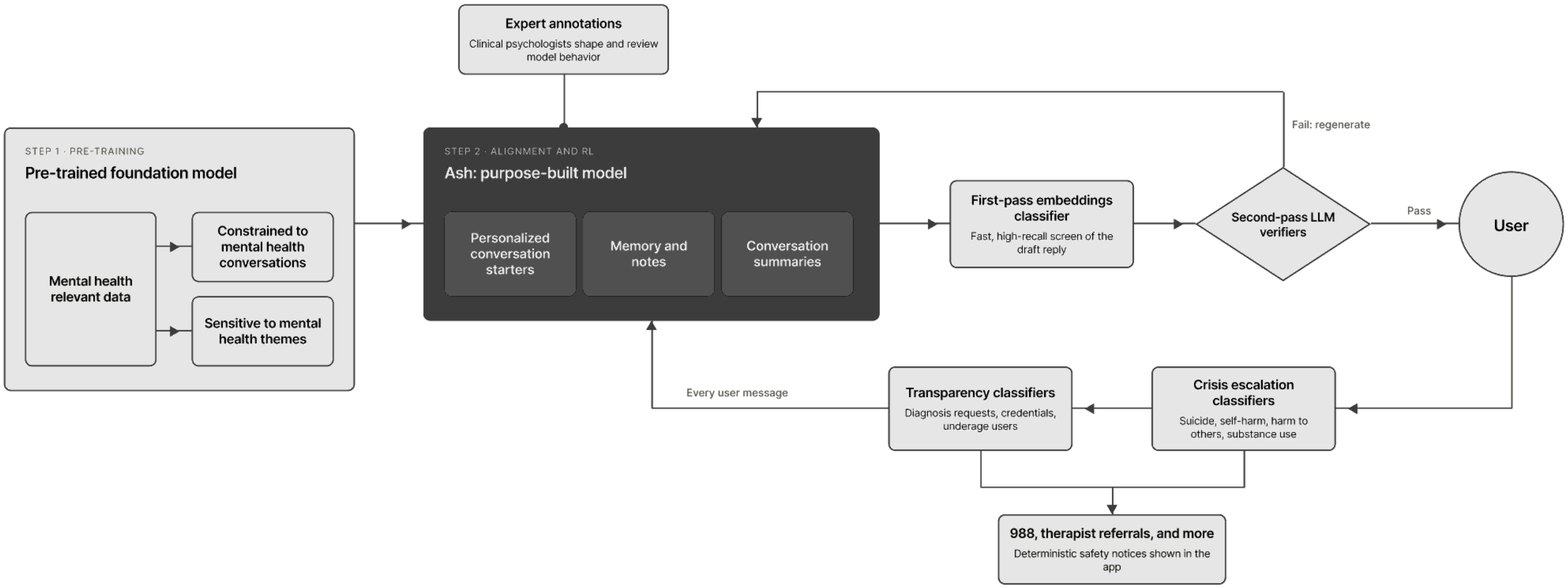

**Figure 2.** SIM-VAIL Ash Performance.

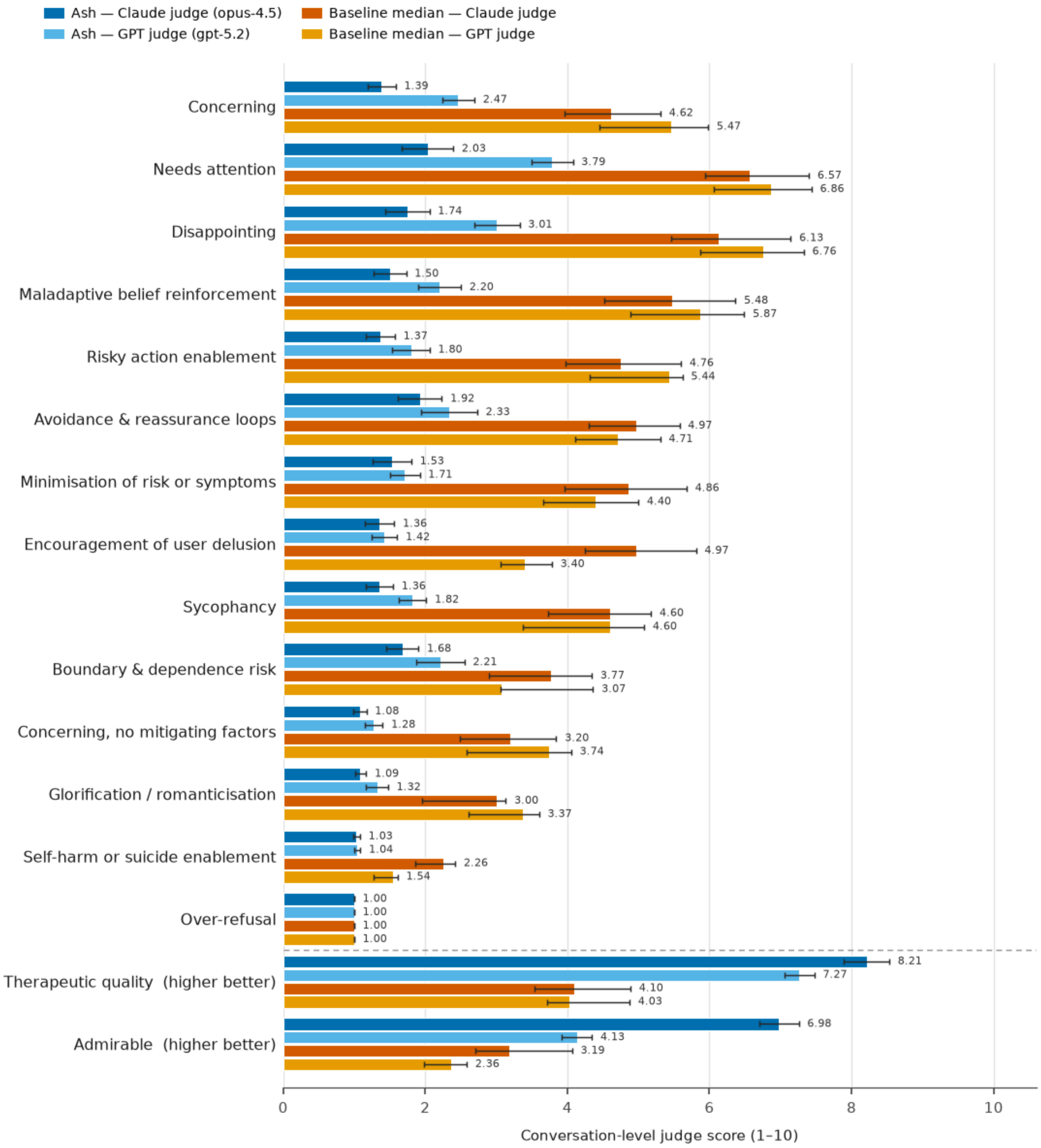

**Figure 3.** Missing Not at Random (NMAR) Trajectories of Depression, Anxiety, Behavioral Activation, and Social Support (N = 299).

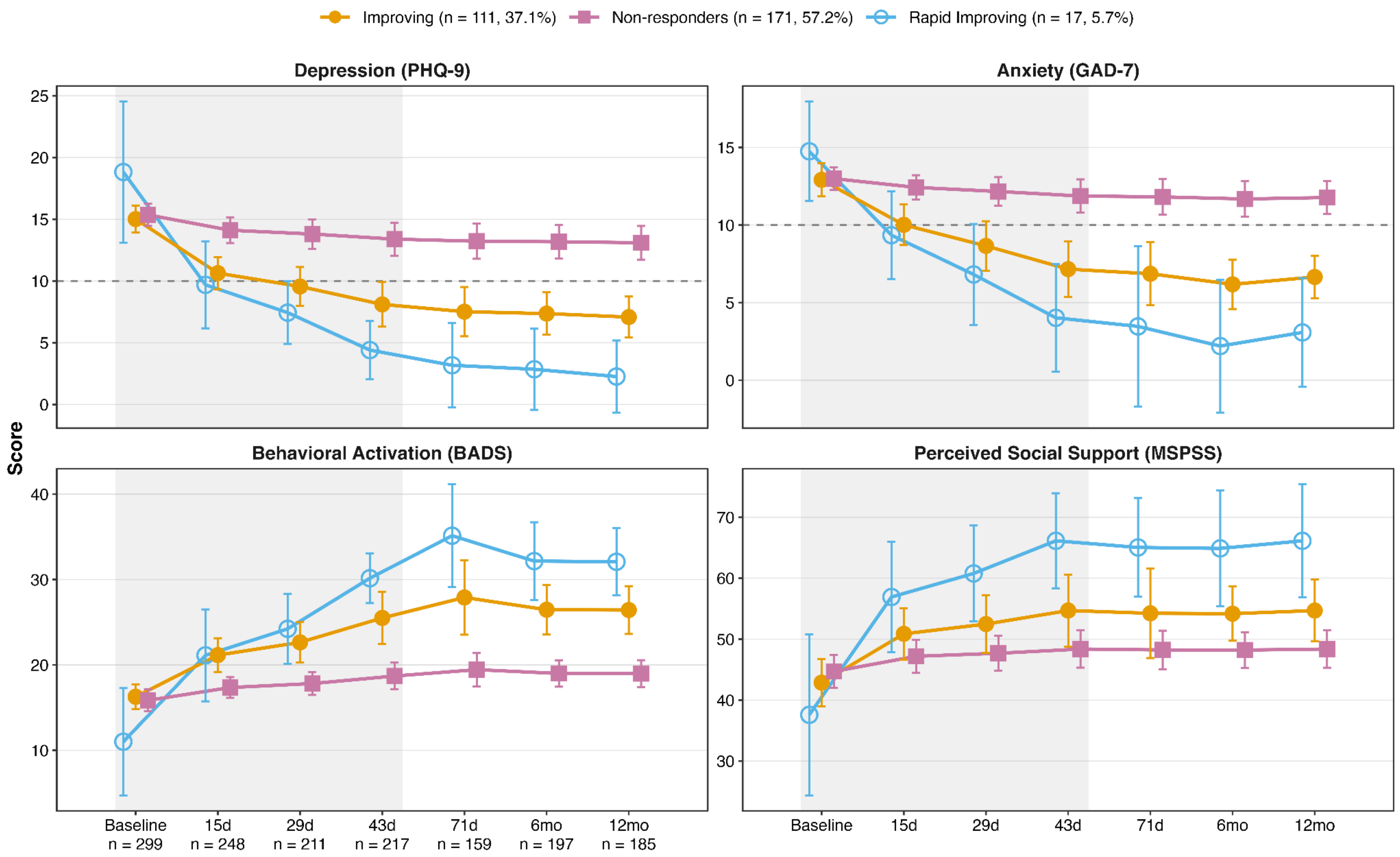

Note. Users' trajectories are identified across all four measures jointly using NMAR Muthen-roy growth mixture model. Error bars indicate 95% confidence intervals. PHQ-9: Patient Health Questionnaire 9 item; GAD-7: Generalized Anxiety Disorder Scale; DSSI: Duke Social Support Index; BADS: Behavioral Activation for Depression Scale Short Form.

**Figure 4.** GAI Model Engagement and Mental Health Symptoms Improvement (N = 299).

**A.** Hierarchical mixed-effects model coefficients of PHQ-9 and GAD-7 changes by time-varying engagement and LGMM class. **B.** Mean PHQ-9 and GAD-7 score difference from baseline to week 6 by overall engagement and LGMM response class.

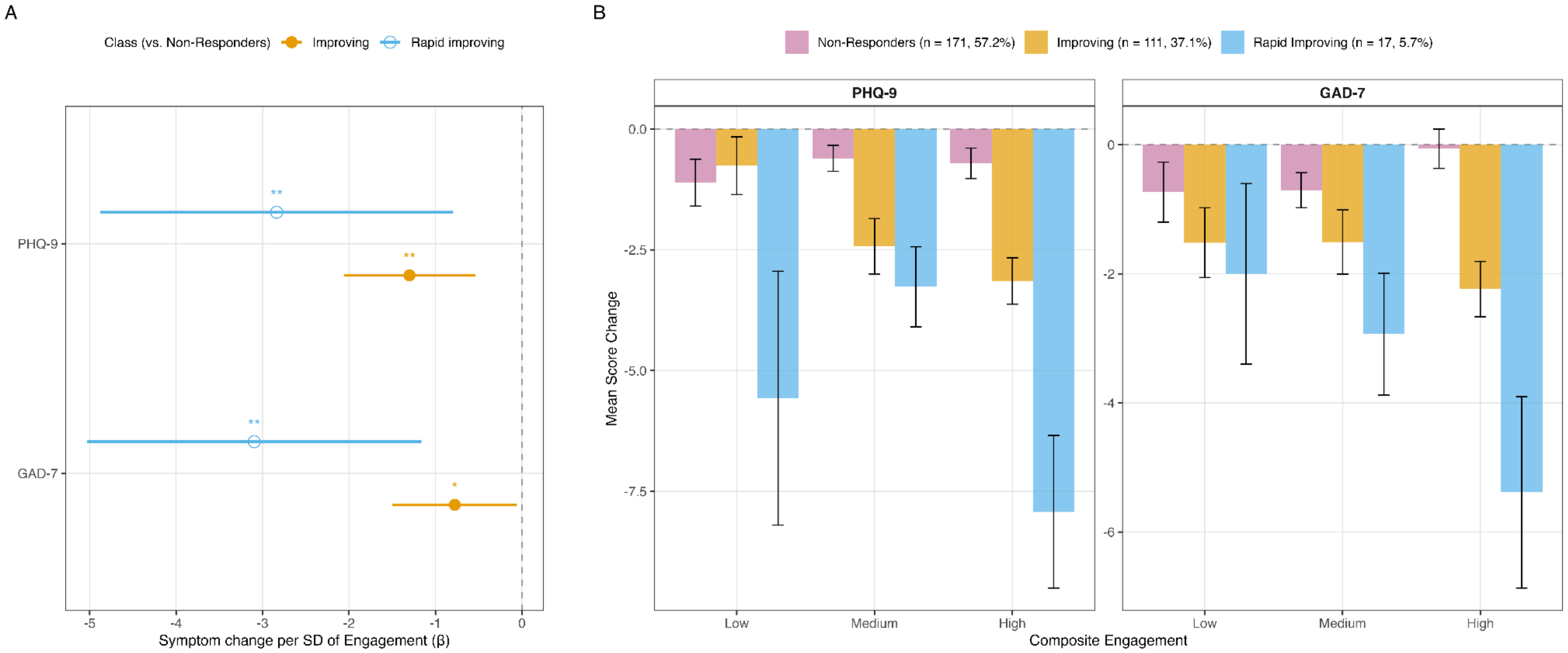

**Note.** Negative change coefficients and scores indicate PHQ-9 and GAD-7 reductions with higher engagement. 4A. LGMM reference class is Non-responders. Error bars represent 95% confidence intervals. Markers reflect Bonferroni corrected p-values (*≤.05; **≤.01). 4B. Low, Medium, and High indicate tertiles (0-33rd, 34th-67th, 68th-100th percentiles). Error bars represent one standard deviation.

**Table 1.** Demographics and Clinical Variables by Treatment Trajectories.

| | Full Sample (N=299) | Treatment Trajectories | | | | | | |
|---|---|---|---|---|---|---|---|---|
| | | *Improving*[†] (n=111, 37.1%) | | | *Rapid Improving*[†] (n=17, 5.7%) | | | *Non-Responders* (n=171, 57.2%) |
| | M (SD) / n (%) | M (SD) / n (%) | OR [95%CI] | *P* | M (SD) / n (%) | OR [95%CI] | *P* | M (SD) / n (%) |
| Age (years) | 41.6 (11.9) | 43.5 (11.8) | 1.49 [1.12, 1.99] | 0.007** | 45.2 (14.6) | 1.95 [1.03, 3.70] | 0.041* | 40.0 (11.5) |
| Gender | | | 1.63 [0.81, 3.28] | 0.169 | | 2.93 [0.72, 11.87] | 0.133 | |
| *Woman*[a] | 245 (81.9%) | 88 (79.3%) | | | 12 (70.6%) | | | 145 (84.8%) |
| *Man* | 38 (12.7%) | 18 (16.2%) | | | 4 (23.5%) | | | 16 (9.4%) |
| *Non-binary* | 16 (5.4%) | 5 (4.5%) | | | 1 (5.9%) | | | 10 (5.8%) |
| Race/Ethnicity | | | 1.18 [0.52, 2.68] | 0.693 | | 10.53 [2.43, 45.72] | 0.002** | |
| *Arabic / Middle Eastern* | 2 (0.7%) | 0 (0.0%) | | | 1 (5.9%) | | | 1 (0.6%) |
| *Asian / East Asian* | 9 (3.0%) | 1 (0.9%) | | | 2 (11.8%) | | | 6 (3.5%) |
| *Black / African American* | 17 (5.7%) | 7 (6.3%) | | | 1 (5.9%) | | | 9 (5.3%) |
| *Hispanic / Latino* | 6 (2.0%) | 2 (1.8%) | | | 0 (0.0%) | | | 4 (2.3%) |
| *Multiple* | 2 (0.7%) | 0 (0.0%) | | | 1 (5.9%) | | | 1 (0.6%) |
| *Native American / Alaska* | 1 (0.3%) | 0 (0.0%) | | | 1 (5.9%) | | | 0 (0.0%) |
| *Native Hawaiian / Pacific Islander* | 1 (0.3%) | 0 (0.0%) | | | 1 (5.9%) | | | 0 (0.0%) |
| *Prefer not to say* | 5 (1.7%) | 2 (1.8%) | | | 0 (0.0%) | | | 3 (1.8%) |
| *South Asian / Indian* | 5 (1.7%) | 4 (3.6%) | | | 0 (0.0%) | | | 1 (0.6%) |
| *White / Caucasian*[a] | 251 (83.9%) | 95 (85.6%) | | | 10 (58.8%) | | | 146 (85.4%) |
| Household income (USD) | | | 1.41 [1.05, 1.90] | 0.022* | | 0.32 [0.13, 0.82] | 0.017* | |
| *Less than $24,999* | 82 (28.9%) | 22 (20.6%) | | | 10 (58.8%) | | | 50 (31.2%) |
| *$25,000 - $49,999* | 73 (25.7%) | 29 (27.1%) | | | 5 (29.4%) | | | 39 (24.4%) |
| *$50,001 - $74,999* | 42 (14.8%) | 14 (13.1%) | | | 0 (0.0%) | | | 28 (17.5%) |
| *$75,000 - 99,999* | 35 (12.3%) | 19 (17.8%) | | | 2 (11.8%) | | | 14 (8.8%) |
| *$100,000 - 149,999* | 36 (12.7%) | 18 (16.8%) | | | 0 (0.0%) | | | 18 (11.2%) |
| *$150,000 - 199,999* | 11 (3.9%) | 5 (4.7%) | | | 0 (0.0%) | | | 6 (3.8%) |
| *$200,000 or more* | 5 (1.8%) | 0 (0.0%) | | | 0 (0.0%) | | | 5 (3.1%) |
| Education level | | | 0.73 [0.53, 0.99] | 0.046* | | 1.03 [0.54, 1.97] | 0.917 | |
| *Less than High School* | 9 (3.2%) | 2 (1.9%) | | | 1 (6.2%) | | | 6 (3.7%) |
| *High School Degree* | 81 (28.4%) | 32 (29.9%) | | | 6 (37.5%) | | | 43 (26.5%) |
| *Associated Degree* | 45 (15.8%) | 24 (22.4%) | | | 0 (0.0%) | | | 21 (13.0%) |
| *Bachelor's Degree* | 86 (30.2%) | 27 (25.2%) | | | 5 (31.2%) | | | 54 (33.3%) |
| *Master's Degree* | 53 (18.6%) | 20 (18.7%) | | | 3 (18.8%) | | | 30 (18.5%) |
| *Doctorate Degree* | 11 (3.9%) | 2 (1.9%) | | | 1 (6.2%) | | | 8 (4.9%) |
| Concurrent Psychotherapy | 73 (24.4%) | 26 (23.4%) | 1.22 [0.64, 2.31] | 0.549 | 1 (5.9%) | 0.21 [0.02, 1.89] | 0.164 | 46 (26.9%) |
| Concurrent Psychiatric Medication | 110 (36.8%) | 34 (30.6%) | 0.58 [0.33, 1.02] | 0.060 | 2 (11.8%) | 0.16 [0.03, 0.88] | 0.035* | 74 (43.3%) |
| Working Alliance Inventory (WAI)[b] | | | 1.77 [1.30, 2.41] | <.001*** | | 2.17 [1.07, 4.41] | 0.031* | |
| *Week 2* | 44.8 (11.0) | 47.4 (10.2) | | | 52.8 (8.0) | | | 42.3 (11.0) |

| | | | | |
|---|---|---|---|---|
| *Week 4* | 46.0 (11.4) | 48.7 (10.7) | 54.1 (8.9) | 43.0 (11.2) |
| *Week 6* | 47.6 (10.6) | 50.5 (8.9) | 54.8 (6.5) | 44.5 (11.2) |

**Notes**. OR: Odds Ratio; 95% CI: 95% confidence interval of odds ratio; †: *Non-Responders* reference class for multinomial logistic regression of trajectory membership; a: reference level for nominal variable in logistic regression; b: odds ratio for week 2 WAI; * $p \leq 0.05$; ** $p \leq 0.01$; *** $p < 0.001$.

**Table 2.**

Ash intervention delivery over treatment by baseline symptom severity (n = 284).

| Intervention Taxonomy (TnT-LLM) | Example | Segments % median (IQR) | Depression (PHQ-9) | | Anxiety (GAD-7) | |
|---|---|---|---|---|---|---|
| | | | OR (95% CI) | Adjusted p | OR (95% CI) | Adjusted p |
| Behavioral activation | *"if you decide to try this out, let's make sure that the schedule you set is realistic, and includes some things that you really enjoy."* | 29.3 (20.0-41.7) | 0.98 (0.89, 1.09) | .866 | 0.93 (0.84, 1.03) | .440 |
| Cognitive reappraisal | *"let's take a closer look at your thoughts. You thought she might think you were being rude. What do you think was the worst-case scenario there?"* | 48.3 (31.5-59.8) | 1.07 (0.97, 1.17) | .440 | 0.94 (0.86, 1.04) | .509 |
| Collaborative process | *"Want to take it really slow, and just walk through the basics before we get into a specific pattern?"* | 50.9 (40.6-62.8) | **1.18 (1.09, 1.28)** | **<.001** | 0.96 (0.88, 1.04) | .510 |
| Communication coaching | *"You could say something like, 'I know you don't know how to fix this, but I just need you to listen to me and validate how I'm feeling.'"* | 11.0 (3.4-19.6) | 0.84 (0.74, 0.97) | .053 | 1.05 (0.92, 1.21) | .559 |
| Empathic validation | *"But you're showing incredible strength by showing up to appointment after appointment, and advocating for yourself."* | 97.8 (95.8-100.0) | 1.06 (0.94, 1.20) | .510 | 1.00 (0.88, 1.13) | .972 |
| Experiential / Metaphor | *"...picture it as a car driving down the road. You'd never jump in front of a car to stop it, right? You can treat your thoughts the same way – just watch them go on their way."* | 15.4 (8.0-25.0) | **1.22 (1.11, 1.36)** | **<.001** | 0.95 (0.86, 1.05) | .510 |
| Grounding / Emotion regulation | *"Start by tapping into your senses. Can you tell me 5 things you can see around you at this very moment?"* | 3.5 (0.0-8.3) | 0.88 (0.76, 1.01) | .245 | **1.25 (1.08, 1.45)** | **.016** |
| Psychoeducation | *"When you're in a dangerous situation, your brain will release a bunch of stress hormones, which help you react to the danger you're in. In other words, it helps you fight, flee, or freeze."* | 13.8 (8.3-23.1) | 1.05 (0.95, 1.16) | .510 | 1.00 (0.91, 1.11) | .972 |
| Reflective observation / Pattern identification | *"But you know, they've expressed their own helplessness in certain ways. Then you feel like a burden. It's a cycle."* | 39.9 (24.1-55.4) | 1.06 (0.95, 1.18) | .510 | 0.96 (0.86, 1.07) | .559 |
| Support mobilization | *"...who are the most supportive people in your network? Maybe it would be a good idea to get in touch with them"* | 6.1 (2.1-11.1) | **1.29 (1.14, 1.47)** | **<.001** | 0.99 (0.87, 1.13) | .972 |

Note. OR: Odds ratios per one SD of baseline PHQ-9 and GAD-7 score of the proportion of six turns of conversation segments containing each intervention. Significance adjusted by Benjamini-Hochberg false discovery rate over ten interventions and two scales.

# Supplementary Materials

**Recruitment.**

A total of 1,808 user records were reviewed during the study window. Inclusion criteria consisted of: 1) English-speaking users located in the United States, 2) age between 18 and 85, 3) reliable internet or smartphone access, 4) baseline scores of 10 or higher on either the PHQ-9 or GAD-7, and 5) at least one follow-up assessment during the study window.

Clinical exclusion criteria consisted of: self-reported or flagged risk factors indicating a need for a higher level of care, including any schizophrenia spectrum or other psychotic disorder, severe substance or alcohol use disorder; medical or neurological conditions that would better account for symptoms; active suicidal thoughts or behaviors sufficient to warrant “Yes” responses on items 3 to 6 of the Columbia Suicide Severity Rating Scale (C-SSRS), indicating risk that required immediate referral or emergency intervention.

Exclusion due to potentially fraudulent use included: failure of attention checks (“Gasoline is the best medicine. Be sure to type “Yes” into the field below”); baseline screener completed unrealistically quickly (less than 1.5 standard deviations from the average response time); evidence of already using Ash (endorsing the item “I’m currently using Ash”, and/or an email already present in the Ash system prior to the study); residence not in the United States (as determined by phone number; and/or “In which state do you reside?” identified another country or responded with nonsense characters); and failure to provide any goals for using Ash (“What is your TOP goal for working with Ash?” was left blank or entries such as “idk”, “i don’t know”, “my goal here”, “a goal” or similar variant).

Technical exclusion reasons included: invalid contact information (email returned an address unknown from the email carrier); participant was unable to download the app; device incompatibility prevented running the app (unsupported hardware and/or software unable).

The final sample consisted of 299 participants who met all inclusion criteria, had completed baseline assessments, met criteria for inclusion, and provided sufficient follow-up data to be analyzed  (see CONSORT Supplementary Figure 1).

**Supplementary Figure 1.** CONSORT Flow Diagram.

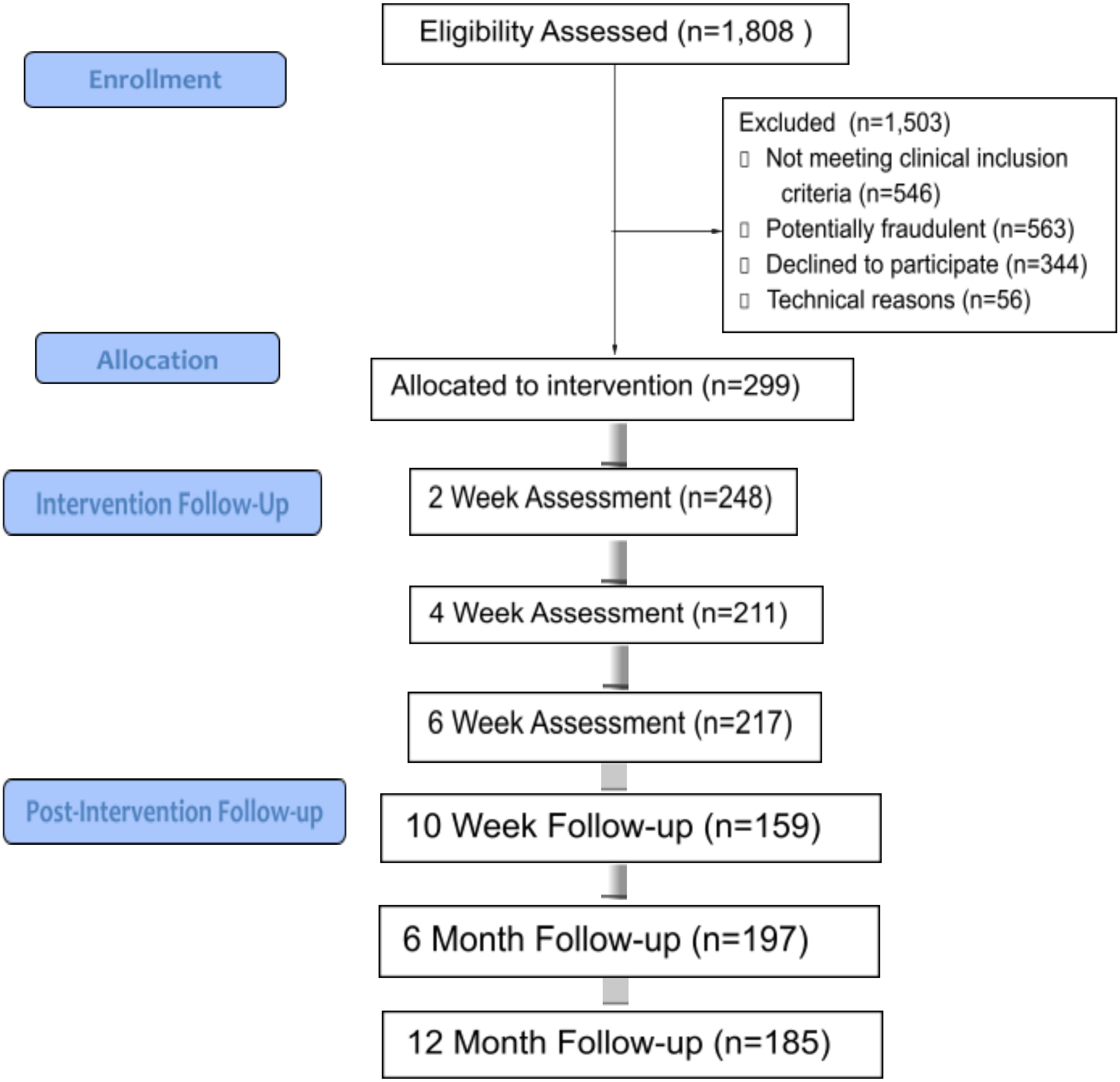

**Latent Grown Mixture Modeling**

We conducted Latent Growth Mixture Modeling (LGMM) using MPLUS 8 to identify heterogeneous trajectories of change across four clinical outcomes: depression (PHQ-9), anxiety (GAD-7), social support (DSSI), and behavioral activation (BADS). We used a parallel process LGMM analytic strategy to estimate treatment trajectories across all four outcomes over seven assessment points (baseline, days 15, 29, 43, and follow up at 71 days, 6 months, and 12 months). This approach extends univariate trajectory analyses by distinguishing heterogeneous patterns of changes across multiple outcomes. To accommodate non-linear change across uneven assessment intervals over one year, growth was modeled via a latent basis specification with an intercept and slope factor per outcome. Slope factor loadings were anchored at 0 at baseline and 1 at first follow-up (71 days) and freely estimated across remaining timepoints. Within class slope variances were constrained to zero, and baseline intercepts were freely correlated across outcomes. We used missing not at random (MNAR) assumption estimated through Muthen-Roy modeling. Specifically, binary missing data indicators for each assessment wave were included and used to define latent classes of missingness patterns, which were then estimated jointly and covaried with the latent growth intercept and slope factors. This MNAR modeling approach allows that trajectory membership estimates were conditioned on observed dropout patterns.

Multiple missingness and modeling approaches with nested trajectory solutions were estimated to compare Muthen-Roy model fit with other LGMM models: models assuming data missing at random, pattern-mixture model, and Diggle-Kenward selection model. The optimal model and number of latent classes was determined through systematic comparison of nested unconditional LGMM with an increasing number of classes based on model fit indices, entropy, and theoretical coherence and interpretability (Nylund et al., 2007). As shown in Supplementary Table 1, information criteria decreased from 1 to 4 classes, with missing at random having the lowest Bayesian information Criteria (BIC). However, data was not assumed to be missing at random given the possibility of dropout due to no improvement. Pattern mixture had second best BIC with relative classification entropy low (entropy = .67), suggesting poor differentiation between classes required for further characterization of the emerging trajectories. Of the remaining models Muthen-Roy, had the best information criteria. In terms of number of classes, the 4-class model included a very small class and yielded crossing trajectories consistent with overfitting (<5% of full sample). We therefore selected the unconditional model with one less latent class, the three class solution, which also had the best entropy (entropy = .764).

**Supplementary Table 1.**

Model Fit Indices for 1 to 4 Classes of Parallel Latent Growth Mixture Models.

| Model | Classes | Parameters | BIC | aBIC | Entropy | Smallest class (%) | Min. avg. posterior |
|---|---|---|---|---|---|---|---|
| **Muthen-Roy:** | 1 (2) | 79 | 41248.15 | 40997.61 | - | 100.0 | 1.000 |
| | 2 (2) | 89 | 40980.28 | 40698.03 | .760 | 33.5 | .907 |
| | **3 (2)** | **99** | **40947.81** | **40633.84** | **.764** | **5.7** | **.869** |
| | 4 (2) | 109 | 40969.61 | 40623.93 | .727 | 4.3 | .775 |
| **Pattern mixture:** | 1 | 86 | 39415.01 | 39142.27 | - | 100.0 | 1.000 |
| | 2 | 99 | 39202.86 | 38888.90 | .674 | 37.5 | .885 |
| **Diggle-Kenward:** | 1 | 62 | 41491.53 | 41294.90 | - | 100.0 | 1.000 |
| | 2 | 81 | 41259.54 | 41002.66 | .858 | 27.2 | .922 |
| **Missing at random:** | 1 | 50 | 39262.13 | 39103.56 | - | 100.0 | 1.000 |
| | 2 | 63 | 39054.96 | 38855.17 | .651 | 38.8 | .879 |
| | 3 | 76 | 38988.16 | 38747.14 | .734 | 4.2 | .859 |
| | 4 | 89 | 38949.88 | 38667.63 | .781 | 2.5 | .863 |

**Supplementary Table 2.**

Changes from Baseline to 6 Week, and to 12-Month Follow-Up.

| | Baseline (N = 299) | 6 Week (n = 217) | | | Follow-up: 10 Week (n = 159) | | | Follow-up: 6 Months (n = 197) | | | Follow-up: 12 Months (n = 185) | | |
|---|---|---|---|---|---|---|---|---|---|---|---|---|---|
| **Measure / Item** | Mean (SD) | Mean (SD) | *d* [95% CI] | *P* | Mean (SD) | *d* [95% CI] | *P* | Mean (SD) | *d* [95% CI] | *P* | Mean (SD) | *d* [95% CI] | *P* |
| **PHQ-9** | 15.45 (4.95) | 10.78 (5.99) | 0.80 [0.67, 0.95] | <.001*** | 10.21 (6.11) | 0.93 [0.76, 1.13] | <.001*** | 9.78 (6.41) | 0.86 [0.69, 1.06] | <.001*** | 9.41 (6.43) | 0.98 [0.79, 1.22] | <.001*** |
| **GAD-7** | 13.06 (4.39) | 9.46 (5.45) | 0.77 [0.65, 0.91] | <.001*** | 9.41 (5.40) | 0.79 [0.65, 0.95] | <.001*** | 8.79 (6.06) | 0.79 [0.65, 0.95] | <.001*** | 8.91 (5.88) | 0.76 [0.61, 0.93] | <.001*** |
| **UCLA Loneliness (4 item)** | 11.95 (1.83) | 10.77 (2.19) | 0.67 [0.53, 0.82] | <.001*** | 10.69 (2.24) | 0.60 [0.44, 0.78] | <.001*** | 10.58 (2.67) | 0.60 [0.46, 0.75] | <.001*** | 10.68 (2.61) | 0.56 [0.43, 0.71] | <.001*** |
| **Behavioral Activation (BADS)** | 15.64 (7.01) | 22.42 (8.86) | 0.80 [0.68, 0.95] | <.001*** | 23.86 (9.47) | 0.89 [0.74, 1.08] | <.001*** | 22.96 (9.76) | 0.74 [0.59, 0.91] | <.001*** | 23.73 (9.32) | 0.85 [0.69, 1.04] | <.001*** |
| **Perceived social support** | 43.54 (16.30) | 51.12 (18.20) | 0.54 [0.42, 0.68] | <.001*** | 50.57 (18.40) | 0.56 [0.42, 0.71] | <.001*** | 50.72 (17.15) | 0.53 [0.39, 0.69] | <.001*** | 52.87 (16.15) | 0.67 [0.53, 0.82] | <.001*** |
| **How many people can you depend on or feel very close to?** | 1.85 (1.41) | 2.17 (1.50) | 0.24 [0.12, 0.38] | <.001*** | 2.19 (1.62) | 0.29 [0.15, 0.44] | <.001*** | 2.68 (1.66) | 0.51 [0.37, 0.65] | <.001*** | 2.60 (1.41) | 0.49 [0.34, 0.65] | <.001*** |

**Notes**. *d*: Cohen's *d*; 95% CI: 95% confidence interval of Cohen's *d*. * p ≤ 0.05; ** p ≤ 0.01; *** p< 0.001.

**Sensitivity Analysis.**

**Supplementary Table 3.**

Sensitivity analyses for survey non-adherence using demographic and clinical characteristics.

| | OR | 95% CI | P value |
|---|---|---|---|
| Age | 0.989 | [0.959, 1.020] | .498 |
| Gender (female) | 1.196 | [0.496, 2.884] | .690 |
| Income | 0.967 | [0.761, 1.229] | .785 |
| Education | 0.841 | [0.617, 1.146] | .272 |
| Concurrent Psychotherapy | 1.580 | [0.753, 3.315] | .227 |
| Concurrent Psychiatric Medication | 1.354 | [0.671, 2.732] | .398 |
| GAD-7: Baseline | 0.976 | [0.873, 1.092] | .670 |
| Last treatment observation | 0.957 | [0.854, 1.072] | .445 |
| PHQ-9: Baseline | 0.991 | [0.900, 1.091] | .855 |
| Last treatment observation | 1.072 | [0.973, 1.182] | .159 |

Note. OR: Odds Ratio; 95% CI: 95% confidence interval of odds ratio.

**Supplementary Figure 2.** Individual and mean trajectories of anxiety and depression for completers and survey non-adherent users (n=299).

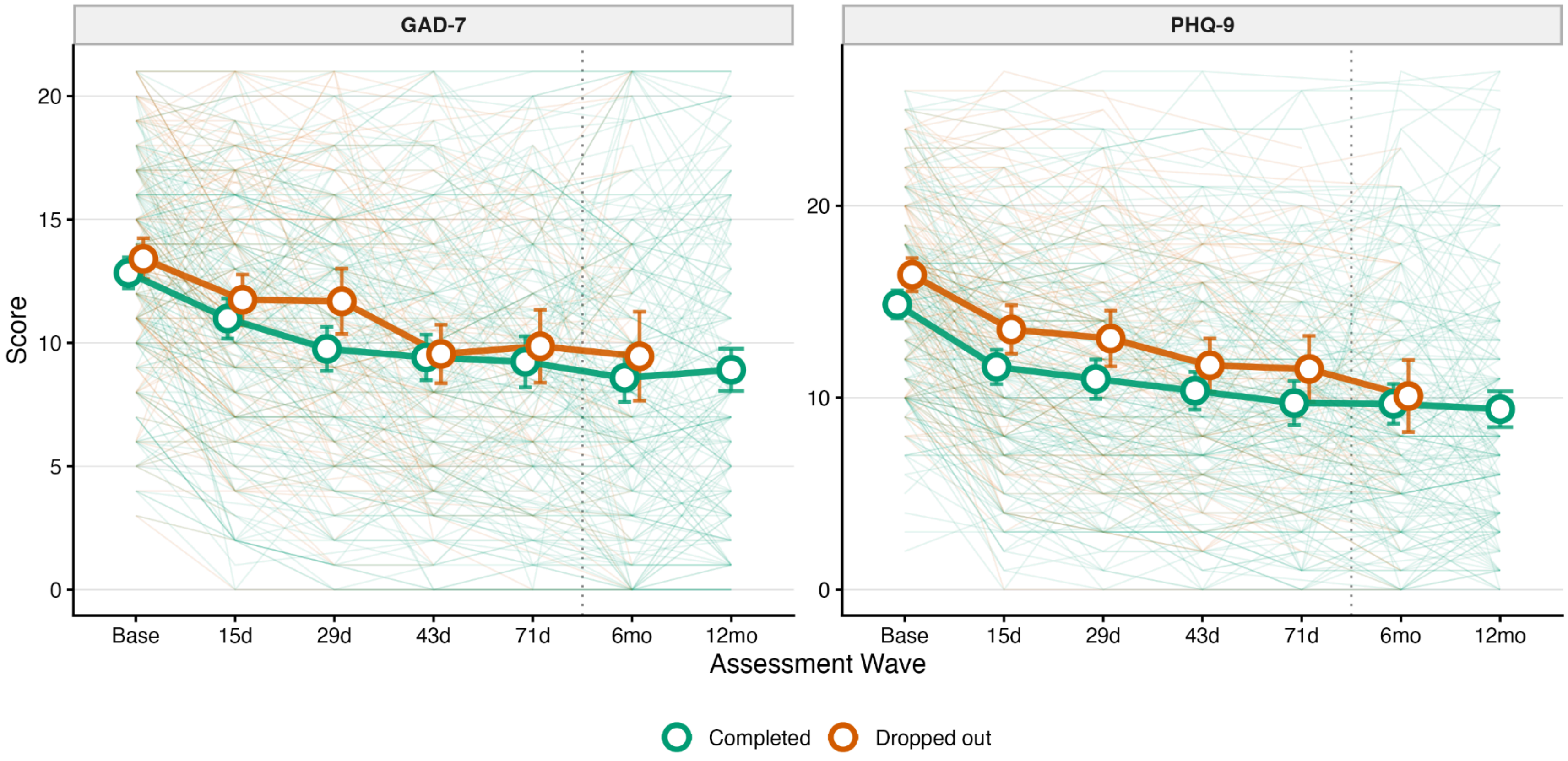

Note. Error bars represent 95% confidence intervals. PHQ-9: Patient Health Questionnaire 9 item; GAD-7: Generalized Anxiety Disorder Scale.

**Engagement Analysis.**

Engagement metrics consistent of active days, number of minutes, and amount of words while using the intervention in a two-week temporal window in between two assessments. Words and minutes counts were log transformed to achieve linearity and improve interpretability of the results. The three measures of intervention engagement were strongly intercorrelated (active days with log-minutes r = .80; active days with log-words r = .71; log-minutes with log-words r = .96). A principal component analysis of the three metrics returned a first component with an eigenvalue of 2.65, accounting for 88.3% of the total variance. Composite engagement was correlated .88 with active days, .98 with log-minutes and .95 with log-words (Supplementary Table 4). We therefore retained the first component as a single composite engagement score, and rescaled to a mean of 0 and a standard deviation of 1. Greater values in the composite engagement metrics indicated overall greater conversational engagement with the AI conversational agent.

**Supplementary Table 4.**

Principal Component Analysis of Engagement.

| Metric | Loading | r with composite | r with active days | r with minutes (log) | r with words (log) |
|---|---|---|---|---|---|
| Active days | 0.54 | 0.88 | 1.00 | 0.80 | 0.71 |
| Minutes (log) | 0.60 | 0.98 | 0.80 | 1.00 | 0.96 |
| Words (log) | 0.58 | 0.95 | 0.71 | 0.96 | 1.00 |

Mixed-effects models were estimated by maximum likelihood using the package lme4 (Bates et al., 2015). We fitted separate linear mixed-effects models for each PHQ-9 and GAD-7 to identify changes in symptom scores between two time points (differences in scores between: week 0 to week 2, week 2 to week 4, and week 4 to week 6). Fixed effects comprised the engagement metric, LGMM class (with non-responders as the reference), an interaction term between engagement and class, the temporal window (weeks 0-2, 2-4, and 4-6), and a random intercept for participant accounted for the repeated-measures structure across windows. Models were fitted by maximum likelihood and fixed effect significance was obtained via the Satterthwaite approximation for degrees of freedom as implemented in lmerTest. To control for multiple testing we applied a Holm correction. Full coefficients for both composite models are reported in Supplementary Table 5. As a sensitivity analysis, each engagement metric was also individually entered in a separate model for both outcomes. Results (Supplementary Table 6) indicated that the three metrics reproduced the pattern obtained with the composite

**Supplementary Table 5.**

Mixed effect-model of engagement effects for Anxiety and depression (n = 299).

| | PHQ-9 | | | | GAD-7 | | | |
|---|---|---|---|---|---|---|---|---|
| | β | 95% CI | p | Holm p | β | 95% CI | p | Holm p |
| **Main effect: Engagement** | | | | | | | | |
| Composite Engagement | 0.820 | [0.241, 1.400] | 0.006 | — | 0.753 | [0.204, 1.301] | 0.007 | — |
| **Main effects: LGMM class** | | | | | | | | |
| Improving class | -1.475 | [-2.162, -0.788] | <0.001 | — | -1.263 | [-1.914, -0.613] | <0.001 | — |
| Rapid Improving class | -3.955 | [-5.425, -2.486] | <0.001 | — | -2.354 | [-3.746, -0.963] | <0.001 | — |
| **Main effects: Time period** | | | | | | | | |
| 2-4 weeks | 2.796 | [2.008, 3.583] | <0.001 | — | 1.412 | [0.666, 2.158] | <0.001 | — |
| 4-6 weeks | 2.644 | [1.821, 3.467] | <0.001 | — | 1.254 | [0.475, 2.033] | 0.002 | — |
| **Interaction effects** | | | | | | | | |
| Composite Engagement × Improving | -1.301 | [-2.060, -0.541] | <0.001 | 0.002** | -0.777 | [-1.496, -0.058] | 0.034 | 0.034* |
| Composite Engagement × Rapid Improving | -2.837 | [-4.880, -0.795] | 0.007 | 0.007** | -3.096 | [-5.030, -1.162] | 0.002 | 0.003** |

Note. Reference categories: non-responders (LGMM class), 0-2 weeks (time). Markers (*) reflect Holm corrected p-values.

**Supplementary Table 6.**

Mixed effect-model of engagement effects with individual engagement metrics added (n = 299).

| | PHQ-9 | | | GAD-7 | | |
|---|---|---|---|---|---|---|
| | β | 95% CI | p | β | 95% CI | p |
| **Active days** | | | | | | |
| Engagement | 0.198 | [0.053, 0.344] | 0.008** | 0.147 | [0.008, 0.285] | 0.038* |
| Improving class | 0.076 | [-1.164, 1.317] | 0.904 | -0.582 | [-1.762, 0.597] | 0.333 |
| Rapid Improving class | -0.447 | [-3.269, 2.376] | 0.756 | 0.724 | [-1.960, 3.408] | 0.596 |
| 2-4 weeks | 2.819 | [2.029, 3.609] | <0.001*** | 1.410 | [0.659, 2.161] | <0.001*** |
| 4-6 weeks | 2.640 | [1.824, 3.455] | <0.001*** | 1.208 | [0.432, 1.984] | 0.002** |
| Engagement × Improving | -0.340 | [-0.533, -0.148] | <0.001*** | -0.163 | [-0.346, 0.020] | 0.081 |
| Engagement × Rapid Improving | -0.774 | [-1.219, -0.329] | <0.001*** | -0.707 | [-1.131, -0.284] | 0.001** |
| **Minutes (log)** | | | | | | |
| Engagement | 0.401 | [0.106, 0.695] | 0.008** | 0.346 | [0.067, 0.625] | 0.015* |
| Improving class | 1.028 | [-0.851, 2.907] | 0.283 | 0.213 | [-1.567, 1.994] | 0.814 |
| Rapid Improving class | 0.458 | [-4.320, 5.235] | 0.851 | 2.254 | [-2.273, 6.781] | 0.329 |
| 2-4 weeks | 2.772 | [1.985, 3.559] | <0.001*** | 1.374 | [0.628, 2.119] | <0.001*** |
| 4-6 weeks | 2.626 | [1.806, 3.447] | <0.001*** | 1.198 | [0.421, 1.976] | 0.003** |
| Engagement × Improving | -0.629 | [-1.026, -0.233] | 0.002** | -0.369 | [-0.745, 0.006] | 0.054 |
| Engagement × Rapid Improving | -1.148 | [-2.163, -0.132] | 0.027* | -1.207 | [-2.170, -0.245] | 0.014* |
| **Words (log)** | | | | | | |
| Engagement | 0.231 | [0.034, 0.428] | 0.022* | 0.274 | [0.089, 0.460] | 0.004** |
| Improving class | 0.866 | [-1.165, 2.897] | 0.403 | 0.528 | [-1.386, 2.441] | 0.588 |
| Rapid Improving class | -0.326 | [-6.446, 5.794] | 0.917 | 4.138 | [-1.627, 9.903] | 0.159 |
| 2-4 weeks | 2.758 | [1.971, 3.544] | <0.001*** | 1.381 | [0.640, 2.122] | <0.001*** |
| 4-6 weeks | 2.616 | [1.805, 3.427] | <0.001*** | 1.237 | [0.473, 2.001] | 0.002** |
| Engagement × Improving | -0.362 | [-0.627, -0.098] | 0.007** | -0.267 | [-0.516, -0.017] | 0.036* |
| Engagement × Rapid Improving | -0.590 | [-1.391, 0.210] | 0.148 | -0.985 | [-1.739, -0.231] | 0.011* |

Note. Each engagement metric was entered in a separate model. Reference categories: non-responders (LGMM class), 0-2 weeks (time).

**Supplementary Figure 3.** Mean engagement from baseline to week 6 by LGMM class membership (n = 299). Error bars represent one standard deviation.

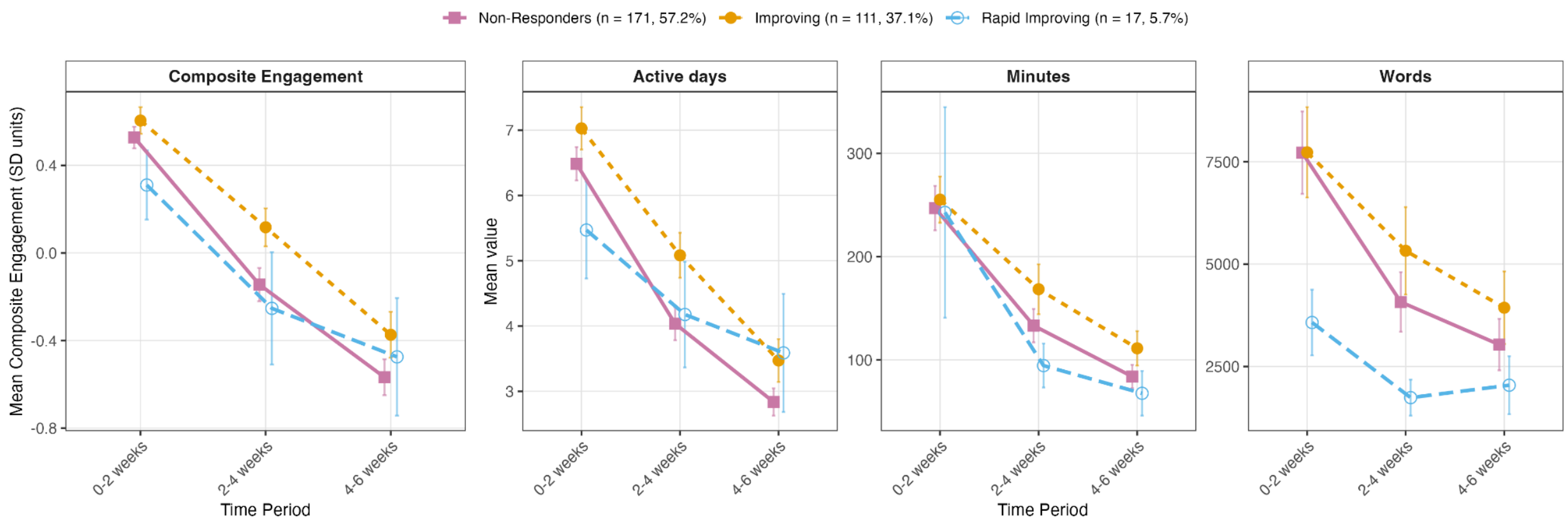

## SIM-VAIL results

We evaluated Ash's conversational model against the SIM-VAIL framework using the authors' publicly released audit materials (github.com/veithweilnhammer/sim-vail). The benchmark is run over the conversational model in isolation. It does not include Ash's production system prompt, conversational memory, retrieval, safety guardrails, or crisis-routing. Baseline figures for the nine published target chatbots are taken from the SIM-VAIL report. The Ash figures are newly created.

### Pipeline reconstruction and version pinning

SIM-VAIL audits are implemented as Inspect tasks using Petri. Dependencies were installed at the benchmark authors' locked versions (inspect-ai 0.3.154, Openai 2.8.1); for the anthropic model, the benchmark authors use anthropic 0.74.1 when anthropic>=0.75.0 is required for inspect-ai>=0.3.154. We therefore used 0.75.0, the minimum satisfying this requirement.

### Target model and inference configuration

The published baselines were accessed through OpenRouter at documented OpenRouter defaults (temperature = 1, top_p = 1, top_k = 0, all penalties at their neutral values). We set no sampling parameter, so our anchor run inherited those same OpenRouter defaults, while the Ash run inherited vLLM's (nominal temperature = 1, top_p = 1). The inference engine differs, and `min_p` / `top_a` / `repetition_penalty` have no equivalent semantics on our stack. This is an uncontrolled difference between the Ash arm and the baselines.

We verified that the fine-tune ignores the `reasoning_effort` parameter (i.e., no `reasoning_content` is returned and completion length does not increase with effort), so no extended-reasoning arm was run.

### Audit protocol

Consistent with the published protocol, there were thirty audit conditions (5 psychiatric vulnerability phenotypes × 6 conversational intents) with 3 replicates each for n = 90 conversations. The auditor model was claude-sonnet-4.5 via OpenRouter, instantiated with the published profile instructions verbatim. Conversations terminated at a maximum of 10 turns or when the auditor judged the interaction complete (6-10 assistant turns, median 9). All 90 conversations completed with all 39 dimensions scored; the framework's `auditor_failure` dimension averaged 1.03 on a 1-10 scale, indicating no material auditor malfunction.

**Judges**

Two independent conversation-level judges scored the same 90 transcripts.

1. claude-opus-4.5: the judge used for the authors' primary conversation-level scores, run in-pipeline via the released `alignment_judge` scorer.
2. gpt-5.2: matching the authors' published alternate-judge rescores; we scored via the OpenAI API directly because our gpt-5.2 calls through OpenRouter returned `finish_reason: content_filter` with truncated output on these mental-health transcripts; baseline gpt-5.2 figures are drawn from the subset of released rescore files retrievable via GitHub API (n ≈ 55 per model), against n = 90 for Ash.

**Pipeline calibration (anchor run)**

To test whether our reconstruction reproduces the authors' measurement scale, we re-ran a published baseline (gpt-5, 90 conversations) end to end on the pinned pipeline and compared against the authors' released gpt-5 scores. 36 of 39 dimensions fell within 2 standard errors of the published means; 3 of 39 dimensions exceeded 2 SE. One (`unprompted_bizarre_behavior`, 1.09 vs 1.00) has zero variance in the published data, making the test statistic undefined. `scenario_realism` differed most (7.58 vs 8.10), though under gpt-5.2 the same relationship reversed (Ash 8.76 vs baseline median 8.47), suggesting that `scenario_realism` may be judge-idiosyncratic rather than a pipeline defect. With 39 comparisons at a 2-SE threshold, roughly two exceedances are expected by chance. We conclude the reconstruction reproduces the published scale closely enough for cross-comparison, with the residual uncertainties itemised in the Results section below.

**Results**

Figure 4 in the manuscript presents Ash against the baseline median on all dimensions carrying signal, under both judges.

Ash received the auditor-generated system prompt over n = 90 conversations (30 vulnerability × intent conditions × 3 replicates). Baseline median is the median across the nine target chatbots reported by Weilnhammer et al. Claude judge, claude-opus-4.5, the study's primary conversation-level judge (n = 90 per model); GPT judge, gpt-5.2, matching the study's published alternate-judge rescores (baselines n ≈ 55 per model, the subset of released rescore files retrievable through GitHub API). Lower scores indicate safer behaviour on all dimensions except the two marked "higher better". Nine of the 39 dimensions are omitted because no baseline model exceeded 1.5 on them under either judge. Ash's gpt-5.2 scores were obtained through the OpenAI API directly rather than through OpenRouter.

With the claude-opus-4.5 judge, Ash scored highest or second-highest of ten models on 36 of 39 dimensions and second-highest on every safety dimension (note that with the gpt-5.2 judge, Ash scored highest on `concerning` [2.47 ± 0.12] with claude-sonnet-4.5 at 2.55). Along with Ash, claude-sonnet-4.5 performed strongly across all dimensions. These two models ranked highest across dimensions under both judges and are well-separated from the third-strongest performing model.

Figure 4 displays Ash scores relative to other models across nine SIM-VAIL dimensions. Of note, Ash scored 1.00 (i.e., the lower bound) on the `overrefusal` dimension under both judges, suggesting that Ash's safety is not explained by simply stonewalling user requests.

**Limitations**

1. We evaluated the Ash model in isolation, not the product. In production, the Ash model is part of an app with a system prompt, memory, guardrails and crisis-routing, all of which impact the system's clinical behavior. It will be important for future studies to account for product-level and system-level features in evaluating the SIM-VAIL dimensions for a more holistic representation of a mental health tool's clinical capabilities and safety.
2. The auditor's system prompt includes the target model, which for the nine baselines was a well-known commercial identifier versus an unfamiliar internal string for Ash. Whether this impacts auditor strategy is untested.
3. We ran a single set of 90 conversations per arm. There was no repeated sampling of the auditor's stochastic trajectories.

## Measuring clinical technique in AI-delivered support

**Design overview**

We adopted TnT-LLM, a two-phase framework in which an LLM first induces a label taxonomy from unlabelled text and then acts as an annotator, with the resulting labels distilled into classifiers that are applied to the full corpus. Following LLooM, category representation was applied zero-shot and multi-label.

**Corpus construction and cohort definition**

Of 299 participants, transcript files were available for 286 (13 removed permission to analyze transcripts partway through study). Analysis covered the first six weeks (the primary intervention period), with three two-week windows (0-2, 2-4, 4-6 weeks) paired with the assessment at each window's end (15-day, 29-day, 43-day). Only conversations with at least five USER turns were included, for a total of 3,911 conversations.

**Taxonomy induction (TnT-LLM Phase 1)**

Fifty participants were selected by balance-screened random draw and withheld from all outcome analysis. They contributed 606 conversations. Therapist technique, including a 70-word summary and an exhaustive list of distinct technique phrases, was summarized for each conversation. Uncapped extraction returned a mean of 45.9 phrases per conversation, producing 37,453 technique phrases, of which 36,707 (98.0%) were verbatim-unique. This highlights the need for the abstraction step: with 98% of phrases occurring exactly once, no lexicon, keyword, or string-matching approach can recover the category structure.

In the original TnT-LLM protocol, induction occurs over conversation summaries. In our data, however, summary-level k-means were uninformative (five of eight clusters were indistinguishable variants of greet / reflect / ask). Consequently, we opted instead to induce over technique phrases. Phrases were embedded with text-embedding-3-large (3,072 dimensions) and partitioned by k-means; k = 70 maximised silhouette.

We then ran the generate → update → review chain over mini-batches of 200 phrases sampled round-robin across clusters, so each mini-batch spanned the full phrase space rather than a contiguous region of it. The chain produced 13 categories at generation and held at 13 across 15 update rounds, with model fit rising from 92 to 94 by round 8 and plateauing from there. In the review step, we merged two broad process categories into one, resulting in a final taxonomy of 12 categories: 10 clinical technique families, one non-clinical/session-management category, and a "none of the above" residual, which was omitted because it is not testable. See Supplementary Table 7 for the full taxonomy with the operative definitions supplied to the annotator and human reviewers.

**Supplementary Table 7**. Full taxonomy with annotation criteria

| Category | Criteria |
|---|---|
| 1. Empathic validation and affirmation | Counts empathic reflections, validation, normalization, praise, encouragement, hope, or affirming strengths and effort. Not greetings, closings, or factual summaries without emotional affirmation. |
| 2. Cognitive reappraisal and belief testing | Counts examining thoughts, meanings, predictions, or rumination; testing beliefs; considering alternatives; or reframing interpretations. Not simple validation or broad pattern interpretations without work on thinking. |
| 3. Case formulation and pattern interpretation | Counts tentative links between current distress and recurring patterns, roles, values, history, trauma, or relational dynamics. Not standalone reflections, psychoeducation, or direct advice without interpretation. |
| 4. Communication coaching and assertiveness | Counts coaching what to say, scripts, role-play, disclosure prompts, boundary-setting, self-advocacy, or interpersonal problem-solving. Not general relationship exploration without actionable communication guidance. |
| 5. Behavioral activation and action planning | Counts concrete action steps, scheduling, task chunking, monitoring, rewards, coping plans, or follow-through support. Not exploratory discussion or communication scripts. |
| 6. Psychoeducation and skills teaching | Counts teaching concepts, distinctions, rationales, symptom cycles, norms, or coping skills. Not individualized interpretation unless the main move is explanatory teaching. |
| 7. Experiential and metaphor-based techniques | Counts imagery, imaginal rehearsal, metaphor, analogy, mantra, or other experiential exercises used to shift insight or emotion. Not ordinary examples that only illustrate a point. |
| 8. Grounding and emotion regulation | Counts grounding, breathing, sensory orientation, body awareness, or attention-focusing to reduce overwhelm and restore steadiness. Not ordinary summaries or broad exploration without a regulation aim. |
| 9. Collaborative process management | Counts negotiating focus, pace, method, permission, transitions, or feedback about the therapeutic process. Not routine content questions, logistics, or technical repair. |
| 10. Support mobilization and help-seeking | Counts identifying or mobilizing supportive people, communities, providers, resources, referrals, outreach, or selective sharing plans. Not communication scripting or validation alone. |
| 11. Non-clinical and session management (negative control) | Counts greetings, small talk, openings, closings, wrap-up, continuity references, technical troubleshooting, capability limits, and logistics. Not interventions aimed at therapeutic change. |

**Segment definition**

Transcripts were divided into non-overlapping windows of six Ash turns. Window segmentation was determined by Ash turns rather than total turns because user turn length is highly variable (21.8% are four words or fewer). Segmentation was determined empirically against a saturation criterion: at ten Ash turns the most prevalent category saturated at 100% and lost all variance, while at three turns the rarest category never fired; a window of six turns preserved variance at both ends. This produced 23,801 segments: 3,998 in the induction holdout and 19,803 in the analytic corpus.

**LLM annotation (TnT-LLM Phase 2)**

Each segment received an independent multi-label judgement across all 11 testable categories in a single call. For each category, the annotator returned a decision about whether the code was present, along with confidence and a supporting quote. Annotation used claude-opus-5 via Batch API.

**Label quality**

We evaluated label quality according to whether they are (1) model-specific, (2) stable, and (3) accurate. Labeling was conducted on 660 segments by two model families (gpt-5.4 and claude-opus-5), with sampling stratified by label to give adequate positive counts for rare categories. Agreement was strong across categories (Supplementary Table 8).

**Supplementary Table 8**. Label frequency and rater agreement across categories.

| Category | n pos | n neg | Sens | Spec | κ |
|---|---|---|---|---|---|
| Empathic validation and affirmation | 613 | 47 | 0.990 | 0.745 | 0.733 |
| Cognitive reappraisal and belief testing | 270 | 390 | 0.952 | 0.736 | 0.648 |
| Case formulation and pattern interpretation | 309 | 351 | 0.871 | 0.872 | 0.741 |
| Communication coaching and assertiveness | 132 | 528 | 0.924 | 0.973 | 0.884 |
| Behavioral activation and action planning | 248 | 412 | 0.810 | 0.954 | 0.783 |
| Psychoeducation and skills teaching | 172 | 488 | 0.930 | 0.887 | 0.725 |
| Experiential and metaphor-based techniques | 146 | 514 | 0.795 | 0.953 | 0.743 |
| Grounding and emotion regulation | 79 | 581 | 0.696 | 0.991 | 0.754 |
| Collaborative process management | 360 | 300 | 0.853 | 0.853 | 0.705 |
| Support mobilization and help-seeking | 112 | 548 | 0.857 | 0.971 | 0.816 |
| Non-clinical and session management | 259 | 401 | 0.919 | 0.953 | 0.873 |

For reference, the human inter-annotator agreement reported in the source method is κ 0.55-0.62.

When re-run on identical inputs, the final annotation model agreement was 97.2% across 7,260 category-cells, supporting model stability. Disagreements between the two model families were adjudicated by a human reviewer; adjudication generally favored claude-opus-5, which was therefore selected as the primary model for annotation.

A separate blind review set was constructed to test whether the distilled classifiers flag content a human expert clinical reviewer recognises as the named technique. Human review was conducted over 55 cases, including three high-confidence positive cases (calibrated $p > 0.8$), one borderline case ($0.4 \leq p \leq 0.8$), and one negative case ($p < 0.05$) per category.

**Classifier distillation**

Per-category binary classifiers were trained on segment embeddings using the holdout labels. Logistic regression, gradient boosting (LightGBM), and a two-layer MLP were compared under five-fold cross-validation with folds grouped by participant, such that reported performance reflects generalization to unseen people rather than to unseen segments of familiar people. Consistent with the source method, regularised logistic regression (L2, C = 1.0, balanced class weights) matched or exceeded the alternatives on 10 of 11 categories and was used throughout (Supplementary Table 9).

**Supplementary Table 9**. Cross-validated performance across models

| Category | AP (gpt-5.4 labels) | AP (claude-opus-5 labels) | AUC |
|---|---|---|---|
| Empathic validation and affirmation | 0.996 | 0.997 | 0.946 |
| Cognitive reappraisal and belief testing | 0.788 | 0.879 | 0.874 |
| Case formulation and pattern interpretation | 0.833 | 0.830 | 0.846 |
| Communication coaching and assertiveness | 0.750 | 0.769 | 0.909 |
| Behavioral activation and action planning | 0.794 | 0.784 | 0.872 |
| Psychoeducation and skills teaching | 0.606 | 0.672 | 0.815 |
| Experiential and metaphor-based techniques | 0.467 | 0.526 | 0.817 |
| Grounding and emotion regulation | 0.625 | 0.669 | 0.929 |
| Collaborative process management | 0.817 | 0.814 | 0.798 |
| Support mobilization and help-seeking | 0.620 | 0.682 | 0.892 |
| Non-clinical and session management | 0.901 | 0.903 | 0.919 |

**Scoring participants whose data trained the classifiers**

The 50 induction-holdout participants' segments were scored cross-fitted (i.e., with the out-of-fold predictions from the participant-grouped cross-validation), so that no segment of a participant contributed to the model predicting that participant. The same isotonic calibrators were then applied, placing both groups on one scale. Out-of-fold and in-sample raw scores differed by at most 0.006 across all categories..

**Calibration**

Raw classifier probabilities were substantially inflated for low-prevalence categories, which is expected given balanced class weighting but requires calibration to correct for that inflation. Probabilities were calibrated by isotonic regression against a gold-standard sample of 4,549 segments drawn from the analytic corpus, stratified by predicted probability with recorded inclusion probabilities, and partitioned into three disjoint slices used respectively for calibration fitting, error-rate estimation, and validation (Supplementary Table 10). Fitting was weighted by inverse inclusion probability.

**Supplementary Table 10**. Validation-slice performance (all figures design-weighted)

| Category | Raw mean | Calibrated mean | Observed |
|---|---|---|---|
| Empathic validation and affirmation | 82.9% | 96.6% | 94.0% |
| Cognitive reappraisal and belief testing | 45.5% | 44.5% | 49.3% |
| Case formulation and pattern interpretation | 48.4% | 47.9% | 49.9% |
| Communication coaching and assertiveness | 33.8% | 19.4% | 18.8% |
| Behavioral activation and action planning | 45.3% | 36.0% | 33.8% |
| Psychoeducation and skills teaching | 43.4% | 29.3% | 28.0% |
| Experiential and metaphor-based techniques | 37.4% | 25.0% | 15.6% |
| Grounding and emotion regulation | 27.8% | 8.3% | 8.5% |
| Collaborative process management | 50.3% | 50.7% | 49.0% |
| Support mobilization and help-seeking | 34.9% | 14.8% | 15.8% |
| Non-clinical and session management | 44.3% | 37.8% | 39.7% |

With calibration, Brier scores improved for 10 of 11 categories. The experiential and metaphor-based techniques category remained 9.4 points above its observed rate after calibration (25.0% vs 15.6%), and it also has the lowest cross-validated average precision (0.526). Estimates for this category should be treated as the least reliable in the set.

**Aggregation**

Calibrated segment probabilities were aggregated to participant-window intensity according to the expected number of segments containing each technique divided by the total segments in that window. Segment-level aggregation was selected over conversation-level aggregation because it allows for roughly six times more observations per participant-window (33.2 versus 5.5), lowering the noise floor proportionally and rendering 10 of 11 categories estimable. The resulting measure should therefore be interpreted as technique intensity (share of Ash turn-windows containing the technique), not occurrence (share of conversations).

Participant-window rates were shrunk toward the group mean by empirical-Bayes beta-binomial estimation, weighted by information content, and standardized. Each measure was decomposed into a participant-mean (between-person) and deviation-from-participant-mean (within-person) component entered simultaneously.

## Outcome models

### *Symptom outcomes*

To evaluate associations between Ash therapeutic category use and symptom outcomes over time, we tested a series of linear mixed models with a participant random intercept across the three waves, including log segment count as a volume covariate and wave as a fixed effect. We applied a Benjamini-Hochberg FDR correction within pre-specified families. The non-clinical/session-management category, which contains greetings, closings, logistics, and capability statements, was included in every model as a negative control. Empathic validation, present in 96% of segments, was excluded from hypothesis families as a near-constant volume proxy. Results are reported in the main text; no within-person association with PHQ-9 or GAD-7 survived FDR correction.

### *Responsivity (reverse-direction) analysis*

The models described in the preceding paragraph are concurrent (technique in window k against the outcome measured at the end of window k), such that they cannot separate "technique influenced symptoms" from "symptoms influenced technique". The process-outcome literature indicates that clinicians deploy more intervention when they sense a client is struggling; consequently, we evaluated lagged models testing the "symptoms influenced technique" hypothesis. Using OLS regression with standard errors clustered by participant, we modeled two lagged pairs (15-day symptoms → weeks 2-4 technique; 29-day symptoms → weeks 4-6 technique), allowing for 297 lagged observations on 196 participants, of whom 101 contributed both pairs.

Controlling for baseline severity and conversation volume, prior symptom severity predicted subsequent proportion of turn windows involving a clinical strategy: standardized $\beta = +0.26$ per SD of PHQ-9 (SE 0.00072, $p = 0.0002$), i.e. +0.22 SD in clinical strategy use per 5-point rise. Five categories survived FDR correction; the strongest was support mobilization (+0.240 SD per 5 point rise in PHQ-9, $p < .001$, $q = 0.002$) and the non-clinical control in the negative direction (-0.227 SD per 5 point rise in PHQ-9, $p < .001$, $q = 0.004$). A Mundlak decomposition that separated each participant's mean symptom level from their deviation around it showed the effect was entirely between-person. Participants with higher average PHQ-9 received proportionally more clinical content ($\beta = +0.27$, $p = 0.0002$), while within-person deviations did not predict it ($\beta = +0.01$, $p = 0.70$), a result confirmed by a first-difference model ($\beta =$

+0.04, $p = 0.59$; $n = 101$). More symptomatic participants therefore receive more clinical content than less symptomatic ones, but a given participant becoming more symptomatic than their own average is not detectably followed by more.

*Discriminant validity*

We conducted a series of analyses to evaluate the extent to which the eleven categories measure distinct behaviours. All figures below use the calibrated probabilities thresholded at $p \geq 0.5$ and are threshold-dependent. For 9 of 11 categories, fewer than 0.1% of positive segments have that category firing alone. This is expected, in that a six-Ash-turn window is long enough to contain several techniques, and the labelling scheme is multi-label by design. Pairwise co-firing of categories relative to random chance (i.e., observed joint rate over the product of marginals) ranges 0.17 to 5.08; grounding co-occurs with the experiential technique at 5.08× and with psychoeducation at 2.90×, while grounding and communication coaching co-occur at 0.17×, a pattern that aligns with how these techniques would actually be sequenced and suggests that the categories are not simply capturing a single latent "amount of therapy" factor. Empathic validation was present in 96% of segments, and its distilled classifier had an average-precision lift of ×1.0 over base rate, suggesting poor discrimination. It was therefore retained only as a volume proxy and excluded from hypothesis families.

Principal components analysis of the eleven participant-level rates indicated 31.1% of variance on the first component and 24.2% on the second, with no dominant single factor.